\documentclass[11pt]{article}
\usepackage{amsmath, amssymb, amsthm, multirow, graphicx, verbatim, subfigure, multicol}

\usepackage[hyperindex,breaklinks]{hyperref}

\usepackage{widecenter}
\usepackage{authordate1-4}
\usepackage{url}
\newcommand{\gammaln}{{\rm gammaln}}
\large
\title{Checking election outcome accuracy \\
Post-election audit sampling}
\author{Kathy Dopp\\
kathy.dopp@gmail.com\\
Comments: 37 pages + 19 pages of appendices \& references, 3 figures, 10 tables\\
Subj-class: stat.AP}
\date{July 10, 2009}
\newcommand{\citetitle}[1]{\emph{#1}}
\begin{document}
\maketitle
\begin{abstract}
    This article advances and improves existing
    post-election audit sampling methodology. Methods for determining post-election audit sampling have been
    the subject of extensive recent research. This article
    \begin{itemize}
    \item provides an overview of post-election audit sampling
    research and compares various approaches to calculating
    post-election audit sample sizes, focusing on risk-lim\-it\-ing
    audits,
    \item discusses fundamental concepts common to all risk-limiting
    post-election audits, presenting new margin error bounds, sampling
    weights and sampling probabilities that improve upon existing approaches
    and work for any size audit unit and for single or multi-winner election
    contests,
    \item provides two new simple formulas for estimating
    post-election audit sample sizes in cases when detailed data,
    expertise, or tools are not available,
    \item summarizes four improved methods for calculating risk-lim\-it\-ing
    election audit sample sizes, showing how to apply precise margin
    error bounds to improve the accuracy and efficacy of existing
    methods, and
    \item discusses sampling mistakes that reduce post-election audit
    effectiveness.
    \end{itemize}

    Adequate post-election audit sampling is crucial because analyzing
    discrepancies found in too-small samples can determine little
    except that the sample size is inadequate. This article is one of
    three articles in a series \citetitle{Checking Election Outcome
    Accuracy}. The other two articles discuss post-election auditing
    procedures and an algorithm for deciding whether to increase the
    sample or to certify the election outcome in response to any
    discrepancies found during a post-election audit.
\end{abstract}

\baselineskip=18pt

\section{Introduction}
In any field the primary purpose of auditing is to detect incorrect results due to unintended innocent or deliberate acts by insiders such as administrators or computer programmers.

This article defines ``post-election audit'' as a check of the accuracy of reported election results done by manually counting all the voter-verifiable paper ballots associated with randomly sampled reported initial vote counts, and checking such additional records as necessary to ensure the integrity of the electoral process.

``Risk-limiting'' post-election audits are election audits that are designed to provide a minimum high probability that incorrect election initial outcomes are detected and corrected before the final certification of election results.

Election winners control budgets and contracts worth millions to trillions of dollars, so this article assumes that election rigging could occur by miscounting the minimum number of initial reported vote counts that could cause an incorrect election outcome (an incorrect winner).

\subsection*{Background}
In 1975, ahead of his time, Roy Saltman proposed conducting post-election
audits using sample sizes that would detect an amount of miscount that could cause an incorrect election outcome, and suggested a formula for estimating the minimum number of miscounted auditable vote counts
 that could cause an incorrect election outcome%
  \cite[appendix B]{Saltman}.\footnote{Saltman calculated the minimum number of minimum miscounted audit units that could cause an incorrect election outcome by using the margin as a percentage of votes cast divided by
 two times a ``maximum level of vote switching that would be undetectable by observation''.}

 Saltman's work and the topic of election auditing was largely neglected until more recently when political scientists, mathematicians, and computer scientists began to recommend that ``rather than relying on ad
 hoc detection and litigation of electoral problems'' that we should ``systematically monitor and audit elections in a preventive fashion''
 \cite{%
      Mebane,%
      Jones,%
      Dill,%
      DoppBaiman,%
      CarterBaker,%
      USGAO%
}.

By 2006, the US League of Women Voters membership voted to recommended post-election auditing and the National Institute of Standards and Technology and The US Election Assistance Commission's Technical Guidelines Development Committee recommended a variety of safeguards for voting systems for the 2007 Voluntary Voting Systems Guidelines \cite[pp. 18--20, 40--41]{Accurate}, \cite{Rivest2006a,Burr2006,NIST,Burr}.

In July 2006 Saltman's formula was re-discover\-ed in a modified form that
 considers the possibility of miscounted over and undervotes%
\footnote{The newer formula handled over and under-votes by
 calculating the margin as a percentage out of the total number of ballots cast, rather than out of votes counted} and
 a numerical computer algorithm was provided for doing these calculations when vote count size varies \cite{%
 	DoppBaiman,%
  DoppStenger%
  }.\footnote{However some authors continued to recommend using
 Saltman's original method based on the number of votes counted
 \cite{%
 		McCarthy,%
 		Norden2007a%
 },
 \cite[Appendix~D]{Hall2008a}} These methods rely on uniform random sampling.

Neff and Wand \cite{Neff,Wand} had showed that the smaller the size of reported vote counts (in number of ballots), the fewer the total number of ballots that need to be audited to achieve the same probability for detecting the level of vote miscount and computer scientists proposed sampling individual ballots to make risk-limiting audits more efficient \cite{Walmsley,Calandrino2007a}.\footnote{In other words, risk-limiting election audits require less work for the same benefit when a larger number of smaller-sized vote counts (audit units) are initially publicly reported and sampled.}  However, most current voting system tabulators are designed to produce reports only of precinct vote counts, thus making it difficult to report and to sample small-sized audit units \cite{Dopp2009c}.

Two groups of computer scientists developed weighted sampling methods
for post-election auditing in order to be able to sample fewer ballots
yet achieve the same probability for detecting incorrect outcomes, by
targeting ballots having more potential for producing margin error
\cite{Calandrino2007a,Aslam2008}. 

In December 2007, a more precise calculation method for post-election
audit sample sizes and sampling weights was developed by using upper
margin error bounds for the just-winning and just-losing candidate
pair
\cite{Dopp2007-8c,Aslam2008,Dopp2008}, \cite[p. 13]{Stark2008a}.%
\footnote{Failure to use accurate within audit unit upper margin error bounds translates to a failure to meet the assumptions that are used to determine the sample. Within audit unit margin error bounds are used in all risk-limiting post-election auditing methods for calculating audit sample sizes and sampling weights, and for analyzing the discrepancies found in the audit.}

Once sampling weights are determined, fair and efficient methods for making random selections for audits have been developed \cite{Cordero2006,Calandrino2007b,Aslam2008,Hall2008b,Rivest2008} although there is some debate among election integrity advocates as to which selection methods are preferred, the more understandable methods such as rolling ten-sided dice, or
computer methods such as pseudo-random number generators that are more efficient and may be more verifiably fair.

\subsection*{Definitions}
An ``audit unit'' or ``auditable vote count'' is defined in this article as a tally of votes that is publicly reported for an election contest. This tally is obtained from a group of one or more ballots that are either:
\begin{itemize}
\item counted at one place and time or
\item counted by one voting device, or
\item cast by voters who live in the same voting precincts or districts.
\end{itemize}
Audit units can be precinct vote counts, electronic voting device counts, or batches or decks of paper ballots. Audit units can be counted by hand or by automatic tabulating equipment where each tally is associated with a number of ballots maintained as a group. An audit unit or auditable vote count may be an individual ballot only if the voting system produces a public report of vote counts on each ballot with humanly readable identifiers for individual ballots and yet preserves ballot privacy.

An ``audit sample size'' is the number of audit units that are randomly drawn for manually counting and comparing with the initial reported audit units.

Under and over-votes are cast ballots eligible to vote in the contest that have no vote counted on them for any candidate.

\subsection*{Assumptions}
\emph{For the purposes of this discussion, it is assumed that:}
\begin{itemize}
	\item effective chain of custody and security procedures are used to prevent and detect any illicit addition, subtraction, substitution, or tampering with ballots and other audit records, and
	\item effective procedures are used during a post-election audit so that
	the manual counts are therefore accurate, and that when
	differences are found, at a minimum, recounts are performed
	until two counts agree --- either the machine and a manual count
	or two manual counts.%
        \footnote{Luther Weeks, in \citetitle{Results of
                  Post-Election Audit of the May 4th Municipal
                  Election} \url{http://www.ctvoterscount.org/?p=2077}
                  suggests that recounts are performed until two
                  counts agree.}
        See \cite{Stevens2007a,DoppStraight,Dopp2009c}.
\end{itemize}

\section{Post-Election Auditing Approaches}
\label{sec:RiskLimitingVsFixedRateElectionAudits}

\begin{table*}[t]
	\caption{Post-Election Auditing Method Comparison}
	\label{tab:AuditMethodComparison}
	\begin{widecenter}
	  \footnotesize
		\begin{tabular}[b]{p{3.5cm}|p{3.5cm}|p{4cm}|p{4cm}}
		\hline
		\textbf{Method} & \textbf{Purpose} & \textbf{Sample Size} & \textbf{Effectiveness} \\ \hline
		\addvspace{2ex} \parbox[b]{4cm}{\textbf{Fixed Rate Audit}} & {\raggedright To ensure that voting machines	are accurate to within a specified tolerance} & {\raggedright A fixed percentage of publicly reported audit units} & {\raggedright Effectiveness at detecting inaccurate initial outcomes ranges widely} \\ \hline
			  \addvspace{2ex} \textbf{Risk-limiting Audit} & {\raggedright To ensure that election outcomes are accurate}  & {\raggedright Varies from one (1) to all audit units, as needed to detect incorrect election outcomes to a desired probability} & {\raggedright Provides roughly equal probability (e.g., at least 95\%) that any incorrect election outcomes are detected} \\ \hline
		\addvspace{2ex} \textbf{Manual Recount} & {\raggedright To ensure that all votes are accurately counted} & {\raggedright 100\% of audit units} & {\raggedright Provides 100\% assurance of detecting incorrect election outcomes} \\ \hline
		\end{tabular}
		\end{widecenter}
\end{table*}

\subsection*{Three approaches to post-election auditing}

The appropriate sample size for conducting post-elect\-ion audits depends on the audit's purpose.  Table~\ref{tab:AuditMethodComparison} compares three approaches to checking the accuracy of reported election results.

\emph{If the purpose of an election audit is to ensure that:}
\begin{itemize}
	\item election outcomes are accurately decided, then \emph{risk-limiting} audits achieve that purpose effectively and efficiently. The risk-limiting post-election audit provides a desired minimum probability that the audit sample will contain one or more miscounted audit units whenever the minimum amount of miscount occurred that would cause an initial incorrect election outcome.
 	\item voting machines have counted election results accurately to within a certain desired tolerance, then a \emph{fixed rate} audit is the solution. Fixed rate audits are commonly used in manufacturing.  Fixed rate audits typically use a larger sample in wide margin election contests, but a smaller sample size in close margin contests than risk-limiting audits.
 	\item every eligible vote is accurately counted, then a \emph{100\% manual} audit or recount is the best solution.
\end{itemize}

\begin{figure*}[tttt] 
	\begin{center}
\includegraphics[width=\textwidth]{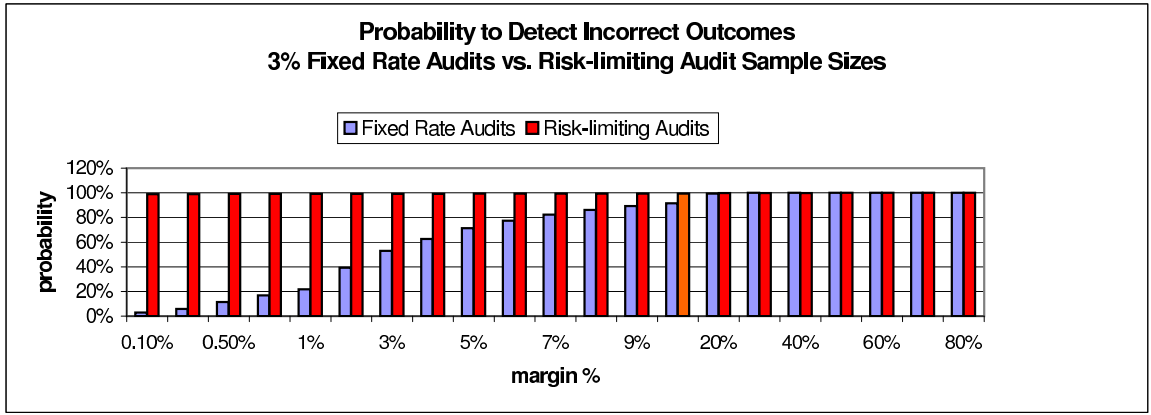}
	\end{center}
	\caption[Detection Probability]{Probability for Detecting Incorrect Election Outcomes}
	\label{fig:Prob2Detect-AuditEst}
\end{figure*}

\begin{figure*}[tttt] 
	\begin{center}
		\includegraphics[width=1.0\textwidth]{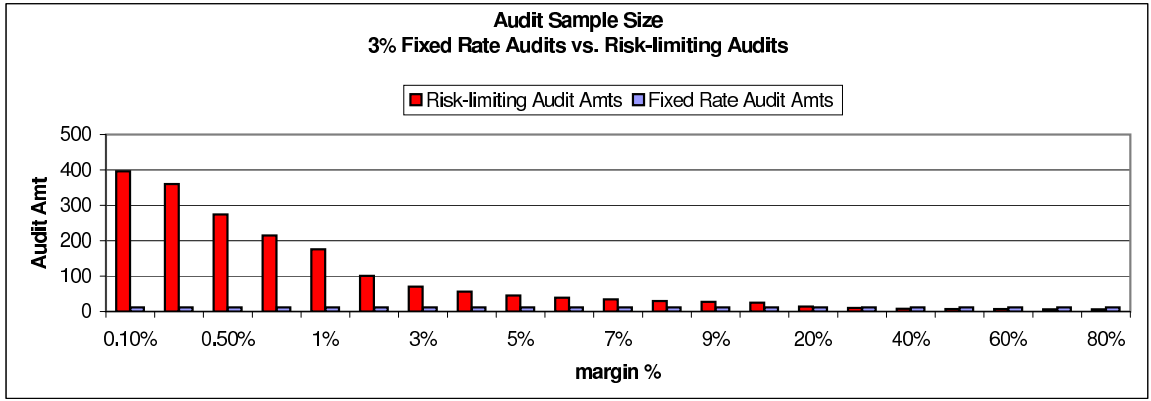}
	\end{center}
	\caption{Sample Size Comparison}
	\label{fig:SampleSize-AuditEst}
\end{figure*}

Figures~\ref{fig:Prob2Detect-AuditEst}~and~\ref{fig:SampleSize-AuditEst}
compare the efficiency and effectiveness of ``fixed rate'' versus
``risk-limiting'' post-election audits using a 500 precinct election
contest with 150,000 total ballots cast and various initial margins.%
\footnote{The audit sample sizes for risk-limiting audits in
Figures~\ref{fig:Prob2Detect-AuditEst}~and~\ref{fig:SampleSize-AuditEst}
are calculated using the uniform estimate method presented later in
this article.}

Figure~\ref{fig:Prob2Detect-AuditEst} shows that 3\% fixed rate audits (blue bars) provide unequal probabilities for detecting the smallest amount of miscount that could cause incorrect outcomes. A 3\% flat rate audit provides very low chance to detect inaccurate outcomes in close contests (less than 10\% probability in some cases), but provides very high minimum probabilities (virtually 100\% in most cases) for detecting miscount that could cause incorrect outcomes in wide-margin contests. On the other hand, risk-limit\-ing audits (red bars in Figure~\ref{fig:Prob2Detect-AuditEst}) provide roughly equal assurance to all candidates and voters that any outcome-changing vote miscount is detected and corrected regardless of the winning margins or the number of precincts.

Figure~\ref{fig:SampleSize-AuditEst} shows how sample sizes for risk-limiting audits increase as winning margins decrease. Thus risk-limiting audits could eliminate the need for automatic recounts because all sufficiently close-margin election contest would automatically receive a 100\% manual count whenever necessary for ensuring that the election outcome were correct.   A fixed 3\% audit (blue bars) samples more than is necessary for ensuring the correctness of wide-margin US House outcomes. In fact, the total overall amount of vote counts that would be audited nationwide for US House contests would be roughly equal for 3\% nationwide fixed rate audits and for 99\% risk-limiting audits.

There are different ways to categorize post-election auditing
approaches based on the sampling approach. Some authors categorize
risk-limiting audits that are designed to detect incorrect election
outcomes in a category called ``Variable (or Adjustable) Rate
Audits'', along with tiered flat rate audits\footnote{An example of a
``tiered'' flat rate audit is the audit proposed by Larry Norden and
Representative Rush Holt, D-NJ that audits 3\%, 5\%, or 10\% of
precincts depending on the margin percentage between the just-winning
and just-losing candidate pair. This particular proposal gives
as low as a 10\% chance of detecting incorrect outcomes when measured
against 2004 US House contest election results.} that do not provide any minimum
probability for detecting incorrect election outcomes
\cite{Norden2007a}, \cite[p. 19]{Norden2007b}, \cite[p. 73]{Hall2008a}. This
article categorizes risk-limiting election audits that \emph{are}
designed to limit the risk of certifying inaccurate election outcomes
in a separate category from election audits that do not limit the risk
of certifying incorrect election outcomes to a desired low
probability.

Despite the recent development of risk-limiting post-election auditing
methods, most States have adopted, and some authors continue to
recommend fixed rate audits designed to detect at most certain levels
of error \cite{Appel,Norden2007a,Norden2007b,Atkeson}.

\section{Risk-limit\-ing Election Audits}
\label{sec:RiskLimitingElectionAuditMethods}

The first risk-limiting post-election audit in the US was conducted in
Cuyahoga County, Ohio in 2006%
\footnote{However, the Cuyahoga County auditors did not correctly analyze the
discrepancies found by the audit based on their sample size design.}
\cite{CollaboratAudit,Dopp2007c}.  Since that time citizen groups in
various States are having limited success convincing election
officials and legislators to implement risk-limiting post-election
audits, and some have been conducted in States such as Colorado and
California \cite{Stark2007,McBurnett,Hall2009}. In January 2009 The League of
Women Voters of the United States endorsed risk-limiting audits,
stating that, ``The number of audit units to audit should be chosen so
as to ensure there is only a small predetermined chance of confirming
an incorrect outcome.'' \cite{LWVUS}

All post-election audit sample sizes and sampling weights are estimates because \emph{all} risk-limiting post-election audit sample sizes and selection weights depend on inputs that are estimates (such as estimates for the number of miscounted audit units that could cause an incorrect election outcome or estimates for the amount of maximum possible within audit unit margin error.) The methods proposed in this article improve upon the accuracy and conservatism of these estimates.

There is more than one method for calculating risk-limiting post-election audit sample sizes. Which method is appropriate depends on the answers to questions such as:

\begin{itemize}
	\item Are initial detailed audit unit and ballot data available for all audit units?
 	\item Is a computer program or spreadsheet available to do the detailed precise calculations?
 	\item Will the random sample be drawn using a uniform probability distribution or by a weighted sampling method?
 	\item Do we need a quick estimate for planning purposes or the precise audit amount that achieves at least the desired minimum probability to an detect incorrect outcome?
\end{itemize}
\begin{table*}[p]
	\caption{This table contains all the variables used in this article to calculate risk-limiting election audit sample sizes and sampling weights.}

\begin{widecenter}
	\footnotesize
	\begin{tabular}[t]{|p{3.3cm}|p{3.4cm}|p{3.6cm}|p{5.5cm}|} \hline
	\multicolumn{4}{|c|}{\textbf{TABLE OF VARIABLES}} \\
	\hline
		\textbf{Variable Name}
		& \textbf{Variable Letter}
		& \textbf{Formula}
		& \textbf{Description} \\
	  \hline
		Ballots cast
		& for each audit unit \(b_i\) and in total $b$
		& {\addvspace{.5ex}} \centering{\(b=\sum_{i}{b_i}\)}
		& the total number of ballots cast that are eligible to vote in the election contest \\
		\hline
		Votes counted for a winning candidate
		& for each audit unit \(w_i\) and in total $w$
		& {\addvspace{.5ex}} \centering{\(w=\sum_{i}{w_i}\)}
		& {\raggedright the total number of votes counted for the winning candidate} \\
		\hline
		Votes counted for the just-losing candidate
		& for each audit unit \(r_i\) and in total $r$
		& {\addvspace{.5ex}} \centering{\(r=\sum_{i}{r_i}\)}
		& the total number of votes counted for the losing candidate who has the most initial votes \\
		\hline
		Votes Counted for Losing Candidates \& Under and Over-votes
		& for each audit unit \(l_i\) and in total $l$
		& {\addvspace{.5ex}} \centering{\(l=\sum_{i}{l_i}\)}
		& the total number of votes counted for any losing candidate plus the total number of ballots with no votes recorded \\
		\hline
		Margin
		& {\addvspace{2ex}} \centering{$M$ between a winning-losing candidate pair}
		& {\addvspace{2ex}} \centering{$w-r$ \\ \addvspace{2ex} \(=\sum_{i=1}^{N}{w_i}-\sum_{i=1}^{N}{r_i} \)}
		& {\raggedright the difference between the number of votes counted for a winning candidate and the number of votes counted for a losing candidate%
		\footnote%
		{In the case of multi-winner contests, Margin is the difference in the number of votes counted for the winning candidate with the least number of votes and the losing candidate with the most number of votes.}}\\
		\hline
		Percentage margin
		& {\addvspace{2ex}} \centering{$m$}
		& {\addvspace{2ex}} \centering{\(M/b\)}
		& {\raggedright the margin divided by the total number of ballots cast eligible to vote in the election contest} \\
		\hline
		Margin error
		& {\addvspace{2ex}} \centering{$e_i$}
		& {\addvspace{.5ex}} \centering{\(e_i=(w_i-r_i)-(w_a-r_a)\)}
		& {\raggedright the difference between the reported initial margins and the audit margins}\\
		\hline
		Margin error upper bound
		& for each audit unit $u_i$ or \(\texttt{error\_bound}(i)\) and for all audit units \(E=\sum_{j=1}^{N}{\texttt{error\_bound}(j)}\)
		& Formula varies, depending on methods and purpose. Total error bound is $E=\sum_{i}u_i$
		& {\raggedright the maximum amount of margin error in \#ballots, within audit units or in total that could reverse an an election outcome} \\
		\hline
		Total number of audit units
		& {\addvspace{2ex}} \centering{$N$}
		&
		& {\raggedright the total number of reported audit units%
                  \footnote{These can be precinct, machine, batch counts or other audit units} in the contest} \\
		\hline
		Maximum level of Undetectability for margin error
		& {\addvspace{2ex}} \centering{$k$ or $MLU$}
		& {\addvspace{2ex}} \centering{a constant $k$: \(0<k<1\)}
		& {\raggedright an assumed maximum rate of margin error that could occur without raising enough suspicion to be detected without an audit%
		\footnote{A value of 40\% to 50\% is suggested. It is assumed that if more than this amount of wrongful margin error occurs within any vote count that comparison of vote counts with comparable results in prior elections or with partisan voter registration or with other data would raise immediate suspicion.  The upper bound of margin error that could contribute to reversing the closest election outcome is the number of ballots cast plus the \#votes counted for the winner minus the number of votes counted for the closest runner-up}
		 }\\
		 \hline
		Confidence probability
		& {\addvspace{2ex}} \centering{$P$}
		& {\addvspace{2ex}} \centering{Suggest \\
		\(0.95 \leq P\leq 1\) \\ See appendix C}
		& {\raggedright the desired minimum probability that the audit sample will detect one or more miscounted audit units if an initial election outcome is incorrect} \\
		\hline
		Audit unit random selection probability
		& {\addvspace{2ex}} \centering{$p_i$}
		& {\addvspace{2ex}} \centering{\(0 \leq p_i\leq 1\) }
		& {\raggedright the probability that an audit unit will be randomly selected} \\
	 \hline
		The number of miscounted audit units
		& {\addvspace{2ex}} \centering{$C$}
		& {\addvspace{2ex}} \centering{methods vary \\ See Table~\ref{tab:PostElectionAuditMethodComparison}}
		& the minimum number of miscounted audit units that could cause an incorrect election outcome \\
		\hline
		Post-election audit sample size
		& {\addvspace{2ex}} \centering{$S$}
		& {\addvspace{2ex}} \centering{methods vary \\ See Table~\ref{tab:PostElectionAuditMethodComparison} }
		& {\raggedright the election audit sample size or number of audit units to manually count and compare with reported results} \\
		\hline
	\end{tabular}

\end{widecenter}
\end{table*}

\subsection{Methods common to risk-limiting post-elect\-ion audits}

\subsubsection*{Maximum level of undetectability}

To reduce chances of detection by a post-election audit, a perpetrator
might miscount the smallest number of total audit units possible to
cause an incorrect election outcome
\cite{Saltman,DoppStenger}. However a perpetrator cannot miscount all
the votes within any one audit unit because if all available votes
were switched to count for the perpetrator's candidate then all voters
who had voted for another candidate would immediately know that the
election results were incorrect. Hence, a smart perpetrator would
miscount at most some maximum rate $k:0<k<1$ of the available margin
error.

Thus we assume a maximum level of undetectability $k$, a maximum rate of margin error, such that if more margin error than $k$ times the upper margin error bound occurs, it would look suspicious and cause immediate action by election officials or by candidates and their supporters.\cite[appendix B]{Saltman}

A risk-limiting audit design assumes that a maximum rate $k$ of the upper margin error bound within audit units or, for individual ballot audit units, that a maximum rate $k$ of certain ballot types overall, could be miscounted in favor of an initial winner without raising immediate suspicion and uses this assumption to estimate a minimum number of miscounted audit units that could cause an incorrect initial election outcome.

When audit units contain multiple ballots, then the larger the assumed level of undetectability as $k\rightarrow 1$, the fewer the number of miscounted audit units it takes to cause an incorrect election outcome; and the larger the audit sample size $S$ must be to ensure that one or more of these potentially miscounted audit units are sampled.%
\footnote{As real-life post-election audits are conducted, more will be learned about what values are most appropriate for the maximum level of undetectability $k$.} Similarly, when individual ballots are the audit units, we still need to multiply the upper margin error bounds for various types of individual ballots times $k:0<k<1$ because we assume that a perpetrator would not target 100\% of all ballots with particular votes on them.

A crucial consequence of making a ``maximum level of undetectability'' assumption when calculating risk-lim\-iting post-election audit sample sizes is that it necessitates allowing candidates or their representatives to select one or more suspicious-look\-ing additional audit units for auditing in addition to randomly selected audit units.

The ``maximum level of undetectability'' is multiplied times the maximum error available in each audit unit to get the most error that it is believed could exist without immediate detection within each audit unit.  However, as seen in the next section, some authors use the actual maximum error, the upper margin error bound, and other authors use an expression such as the number of votes cast that is an inaccurate measurement for the total possible error.

The use of an assumed ``maximum level of undetectability'' necessitates a procedure of allowing losing candidates to select discretionary audit units to be manually audited at the same time as the randomly selected audit units.  The necessity of auditing additional discretionary audit units is discussed in section \ref{sec:OtherConsiderations} in this article.

One proposed approach  is to use 1 or 100\% for the maximum level of
  undetectability in sample size
  calculations \cite{Stark2008a,Stark2008b,Stark2008c,Hall2009}. This would
  normally result in unnecessarily conservative (large) sample sizes if any of the methods suggested in this article were used, but the maximum level of undetectability is in effect canceled from both sides of an inequality involving the ratios of different within precinct upper margin error bound measures than those recommended herein \cite[p. 6--10]{Stark2009a} \cite[p. 4]{Stark2009b}.  This cancellation is similar to how the maximum level of undetectability cancels when calculating the sampling weights for the probability proportional to margin error bound with replacement (PPMEBWR) method described below.  Assuming a 100\% level of undetectability is of course unreasonable because some voters and candidates would immediately notice the fact that there were zero votes cast in their precincts for their candidates. A perpetrator can not expect to steal 100\% of available target votes for his candidate and not be noticed.

\subsubsection*{Upper margin error bounds}

Why are within audit unit upper margin error bounds important?

Within audit unit upper margin error bounds are a crucial input to all calculation methods for determining post-election audit sample sizes and sampling weights.

Aslam, Popa, and Rivest first derived precise within audit unit upper margin error bounds for particular winning-losing candidate pairs in an intermediate calculation, but recommended Saltman's earlier method of multiplying the maximum level of undetectability, $s=20\%$, times two ($2$) times the number of votes cast, $v$, to approximate the maximum undetectable margin error for their sample size calculations.

At about the same time Dopp derived the within audit unit upper margin error bounds, $b+w-r$ (the number of ballots plus the margin in votes), and applied it to improve the accuracy of post-election auditing sample size calculations in place of her original recommendation of $2sb$ where $b$ is the number of ballots cast.  Later Stark also recommended using within audit unit upper margin error bounds, but incorrectly took the maximum of normalized upper margin bounds of all winning-losing candidate pairs and negated the use of the upper margin error bounds by employing an arbitrary small level of ``acceptable error $t$'' when calculating sample sizes and analyzing discrepancies.

For risk-limiting audits, when audit units are larger than one ballot,
most authors continue to use or recommend using the less precise
expression $2sv$ for approximating maximum undetectable within audit
unit margin error
\cite[p.~5]{Calandrino2007a}, \cite[p.~16]{Aslam2008}, \cite[p.~5 of
Appendix~B]{Saltman}, \cite[p.~6 and Appendix~B]{McCarthy},
\cite[pp.~6--10, 15]{Stanislevic}, \cite[Appendix~B]{Norden2007b},
\cite[p.~153, Appendix~D]{Hall2008a}, \cite[pp.~11, 28--30]{LWVUS},
\cite{McBurnett}.

Using less precise margin error bounds can produce insufficient sample sizes and, if used for sampling weights, can cause the audits to unfairly favor some candidates over others.  For example, using the expression $2sv$ to approximate margin error:
\begin{enumerate}
	\item ignores the partisanship of within audit unit vote counts. The partisanship of the vote counts affect the amount of maximum margin error that can exist between specific winning-losing candidate pairs and between all winning-losing pairs,
	\item calculates an impossible amount of error when certain vote shares occur --- more than the possible margin error that is available to contribute to causing an incorrect outcome --- in cases when just-losing candidate vote share is high,
	\item does not account for the unequal amounts of margin error that results from different causes such as shifting votes from a winner to a loser or vice-versa, shifting votes between two different losing candidates or between different winning candidates, and not counting votes for a losing or winning two candidates. Each miscounted vote may produce a vote margin error of -2, -1, 0, 1, or 2 in the initial margin for a particular winning-losing candidate pair. There is no way to express these variations precisely in terms of votes or ballots cast using an expressions like $2sv$ or $2sb$ without making awkward assumptions about the relative proportions of each type of error (such as vote-switching between two winners, a winner and a loser, or two losers or simply not counting votes);
	\item by using votes cast, $v$, fails to consider miscounted under or over-votes (although that particular problem could be corrected by using ballots cast, $b$);
	\item significantly understates the possible margin error in most cases because actual upper margin error can be as high as 200\% of the number of ballots  (Eg. 40\% times $ 0 \leq s \leq 1$ times the number of cast votes significantly under-states 40\% times the amount of possible within audit unit margin error); and
\end{enumerate}

Thus using $2sv$ or $2sb$ as an estimate of maximum error in close margin contests underestimates the sample size, or equivalently over-states the probability for detecting the minimum level of vote miscount that could cause an incorrect election outcome.%
\footnote{Dopp compares and contrasts the use of votes cast versus using the actual within audit unit upper margin error bounds to calculate audit sample sizes in \cite{Dopp2007c}.}

This article presents two types of precise within audit unit upper margin error bounds that are should be used when calculating risk-limiting post-election audit sample sizes and sampling weights:
\begin{enumerate}
	\item upper margin error bounds for a specific winning-losing candidate pair, and
	\item upper margin error bounds for the error that can occur between any candidate pair
\end{enumerate}

Figure~\ref{fig:Error-2svBound} in Appendix~D graphically compares the $2sv$ margin error measure with the actual upper error bounds.

\subsubsection*{Upper margin error bounds for specific winning-losing candidate pairs}


The margin error between any winning-losing candidate pair is defined as the signed difference between their initial reported margin and their margin found during a 100\% manual audit.
Within each initial audit unit $i$, let
\begin{align*}
  b_i &= \text{~the number of ballots cast,}\\
  w_{ij} &= \text{~the number of initial votes for winning candidate,}\\
  l_{ij} &= \text{~the number of initial votes for losing candidate,}\\
  w_{aj} &= \text{~the number of audit votes for winning candidate,}\\
  l_{aj} &= \text{~the number of audit votes for losing candidate.}
\end{align*}
Then the margin error within each audit unit $i$ found during the audit between winning-losing candidate pair $j$ is the difference between the initial margin and the audit margin:

\begin{equation*}
e_{ij}=(w_{ij}-l_{ij})-(w_{aj}-l_{aj})
\label{eq:marginErr}
\end{equation*}

 and in all cases
\begin{equation*}
(w_{ij}-l_{ij})-(w_{aj}-l_{aj})\leq{b_i+w_{ij}-l_{ij}}
\end{equation*}
because $-(w_{aj}-l_{aj})\leq{b_i}$ \\
Therefore, authors agree \cite{Dopp2007-8a,Dopp2007-8c,Aslam2008,Stark2008a} that the maximum possible within audit unit initial margin error that could occur between any win\-ning-los\-ing candidate pair $j$ for each audit unit $1\leq i \leq n$ is

\begin{equation}
u_{ij}=b_i+w_{ij}-l_{ij}
\label{eq:meb}
\end{equation}

Note that, in the case of a multi-winner contest and individual ballot audit units, the expression for the upper margin error that any vote on the ballot could contribute to a winning-losing candidate pair ($w_{ij}$ and $l_{ij}$) reduces to:
\begin{align*}
b_i+w_{ij}-l_{ij}&=2 \text{~if a vote is for winning candidate $j$,} \\
&=1 \text{~if a vote is for another winning candidate,}\\
&=0 \text{~if a vote is for losing candidate $j$,}\\
&=1 \text{~if a vote is for another losing candidate,} \\
&=1 \text{~if a vote is an over or under-vote}
\end{align*}

The upper margin error bound for that ballot is also zero (0) when all losing candidates receive votes on the ballot.~\cite[p. 7]{Calandrino2007a}

The audit unit upper margin error bound for winning-losing candidate pair $j$ can be written as a percentage by diving by the number of ballots cast, resulting in the expression
\begin{equation*}
(1+(w_j-l_j)/b_i)
\end{equation*}.

For a simple example of the overall upper margin error, if the initial election results shows the initial winner has 100 votes and the initial runner-up has 0 votes with no under-votes or other candidates, but a full manual recount shows that the initial runner-up really had 100 votes and the initial winner had 0 votes, then the total margin error is $200=100+100-0$ votes or 200\%.

\begin{description}
	\item[Example] If the winner has 51\% of the reported ballots
	cast and the runner-up has 48\%, then the reported margin is
	3\%.  For the reported winner to be incorrect there must be at
	least 3\% margin error plus one vote.  What is the minimum
	number of corrupt vote counts that could cause 3\% or more
	margin error and thus result in an incorrectly reported
	election winner?  The total possible percentage margin error
	in this example contest is 103\% if all votes not counted for
	the runner-up should actually have been counted for the
	runner-up, so that the vote share of the runner-up should have
	been 100\% with 0\% for all other candidates.  The upper
	margin error bound is thus found by taking the actual margin
	minus the reported margin between the winner and the runner-up
	or $100\% - (-3\%) = 103\%$.
\end{description}

\begin{table*}[tttt] 
	\label{tab:PerVoteMarginErrorAB}
	\begin{widecenter}
	  \footnotesize
		\begin{tabular}[b]{|p{3cm}|c|c|c|c|}
		\hline
		\multicolumn{5}{|c|}{\textbf{Per Vote Miscount-Caused Margin Error for Candidate Pair A \& B}} \\
		\hline
		\textbf{Vote initially reported \& counted for} & \multicolumn{4}{|c|}{\textbf{Vote actually cast for }} \\
		\cline{2-5}
		 & {\textbf{candidate A}} & {\textbf{candidate B}} & {\textbf{candidate C}} & {\textbf{under or over\-vote}} \\
		 \hline
	\textbf{Initial winning candidate A}	&	  \textbf{x} & {\textbf{2}}  & {\raggedright 1} & {\raggedright 1} \\
		\hline
	\textbf{Initial losing candidate B}	& \textbf{--2} & {\textbf{x}} & {\textbf{--1}} & {\textbf{--1}} \\
		\hline
		\textbf{Initial losing candidate C}	& \textbf{--1} & {\textbf{1}} & {\textbf{x}} & {\textbf{0}} \\
		\hline
		\textbf{Initial under or over-vote}	& \textbf{--1} & {\raggedright 1} & {\raggedright 0} & {\raggedright x} \\
		\hline
		\end{tabular}
		\end{widecenter}
		\caption{This table shows all possible error
                        values that one miscounted vote could cause for the
                        margin between a winning candidate A and a losing
                        candidate B due to vote mis-allocation in a one-winner
                        (one-seat) contest with three candidates A, B, and
                        C\@. The margin error is 2 when a vote is initially
                        counted incorrectly for the winning candidate A that
                        should have been counted for candidate B; is 1
                        when a vote is initially reported for winning
                        candidate A that should have gone to another losing
                        candidate or should have been an under or over-vote;
                        and is 1 when a vote is initially reported for another
                        losing candidate or as an under or over-vote that
                        should have been counted for candidate B.}
\end{table*}

\subsubsection*{Upper margin error bounds for \emph{all} winning-losing candidate pairs}
Notice that one initial incorrectly recorded vote can contribute
margin errors of $-2, -1, 0, 1$ or $2$ votes (See
figure~\ref{tab:PerVoteMarginErrorAB}), so that the maximum margin
error that any individual vote for a winning candidate can contribute
is $2$ votes. One initial incorrect vote for an initial losing
candidate or an under or over-vote can contribute margin errors of
$-2, -1, 0$, or $+1$ votes, so that the maximum margin error that an
initial vote for a losing candidate can contribute is $1$ vote.
Therefore the upper margin error bound for all winning-losing
candidate pairs within each audit unit $i$ is

\begin{equation}
u_i= 2\sum_i{w_i}+\sum_i{l_i}
\label{eq:weights1}
\end{equation}

where $l_i$ is the number of total votes for any losing candidates plus the number of total under or over-votes (cast ballots eligible to vote in the contest that have no vote counted for any candidate) and $w_i$ is the number of total votes for any winning candidate within audit unit $i$.

Note that, in the case of a single-winner contest, when using individual ballot audit units, the expression for the maximum margin error bound that could contribute to a losing candidate becoming the actual winner for each winning-losing candidate pair reduces to:
\begin{align*}
2\sum_i{w_i}+\sum_i{l_i}&=2 \text{~if the vote is for the winning candidate,} \\
&=1 \text{~if the vote is for a losing candidate,} \\
&=1 \text{~if the vote is an over or under-vote}
\end{align*}

Note that an equivalent, perhaps simpler, expression for the upper margin error bound for \emph{all} winning-losing candidate pairs is simply
\begin{equation}
u_i= b_i + \sum_i{w_i}
\label{eq:weights2a}
\end{equation}
where $b_i$ is the number of total ballots cast in audit unit $i$.  The formula is equivalent because the number of ballots includes votes for all winning and losing candidates and under and over-votes, and  $\sum_i{w_i}$ doubles the number of votes for the winning candidates.

Calandrino, Halderman, and Felten \cite{Calandrino2007a} point out that there is an exception to the above rule. There must be at least one losing candidate not on the ballot in order for a ballot to contribute any margin error that could cause an incorrect election outcome.  Therefore the upper margin error bound is zero for a ballot in the case that all losing candidates have votes on a particular ballot because such a ballot can only contribute negative margin error towards causing an incorrect outcome.

\begin{table*}[htbp] 
	\label{tab:EgDataMEBs1}
	\begin{widecenter}
	  \footnotesize
		\begin{tabular}[b]{|p{2.5cm}|c|c|c|c|p{2.5cm}|p{2.5cm}|}
		\hline
		\multicolumn{7}{|c|}{\textbf{Example Upper Margin Error Bounds in a Wide Margin Contest}} \\
		\hline
		\multicolumn{5}{|c|}{\textbf{2004 Utah State Rep, Dist 3 Vote Counts}} & \multicolumn{2}{|c|}{\textbf{Upper Margin Error Bnds}} \\
		\hline
		 {\textbf{Precinct}} & {\textbf{\#Ballots}} & {\textbf{Buttars}} & {\textbf{Hurtson}} & {\textbf{Elwell}} & {\textbf{Just-winning/just-losing candidate margin error bnds}} & {\textbf{All candidate pair margin error bnds}} \\
		 \hline
	\textbf{Totals}	&	 \textbf{13,495} & {\textbf{9,614}}  & {\textbf{2,930}} & {\textbf{236}} & {\textbf{20,179}} & {\textbf{23,109}} \\
		\hline
	\textbf{SMI1}	& \textbf{871} & {\textbf{638}} & {\textbf{185}} & {\textbf{11}} & {\textbf{1,324}} & {\textbf{1,509}}\\
		\hline
		\textbf{HYD2}	& \textbf{996} & {\textbf{599}} & {\textbf{340}} & {\textbf{16}} & {\textbf{1,255}} & {\textbf{1,595}}\\
		\hline
		\textbf{SMI2}	& \textbf{819} & {\textbf{599}} & {\textbf{165}} & {\textbf{15}} & {\textbf{1,253}} & {\textbf{1,418}}\\
		\hline
		\textbf{NL04}	& \textbf{864} & {\textbf{595}} & {\textbf{222}} & {\textbf{20}} & {\textbf{1,237}} & {\textbf{1,459}}\\
		\hline
		\textbf{NL03}	& \textbf{754} & {\textbf{553}} & {\textbf{144}} & {\textbf{17}} & {\textbf{1,163}} & {\textbf{1,307}}\\
		\hline
		\textbf{SMI5}	& \textbf{736} & {\textbf{543}} & {\textbf{152}} & {\textbf{14}} & {\textbf{1,127}} & {\textbf{1,279}}\\
		\hline
		\textbf{NLO1}	& \textbf{787} & {\textbf{512}} & {\textbf{208}} & {\textbf{17}} & {\textbf{1,091}} & {\textbf{1,299}}\\
		\hline
		\textbf{SMI4}	& \textbf{700} & {\textbf{510}} & {\textbf{144}} & {\textbf{10}} & {\textbf{1,066}} & {\textbf{1,210}}\\
		\hline
		\textbf{LO17}	& \textbf{699} & {\textbf{488}} & {\textbf{132}} & {\textbf{13}} & {\textbf{1,055}} & {\textbf{1,187}}\\
		\hline
		\textbf{NL02}	& \textbf{633} & {\textbf{465}} & {\textbf{108}} & {\textbf{19}} & {\textbf{990}} & {\textbf{1,098}}\\
		\hline
		\textbf{LO25}	& \textbf{600} & {\textbf{439}} & {\textbf{93}} & {\textbf{10}} & {\textbf{946}} & {\textbf{1,039}}\\
		\hline
		\textbf{LO04}	& \textbf{582} & {\textbf{389}} & {\textbf{138}} & {\textbf{11}} & {\textbf{833}} & {\textbf{971}}\\
		\hline
		\textbf{RCH1}	& \textbf{507} & {\textbf{390}} & {\textbf{89}} & {\textbf{13}} & {\textbf{808}} & {\textbf{897}}\\
		\hline
		\textbf{RCH2}	& \textbf{503} & {\textbf{389}} & {\textbf{87}} & {\textbf{9}} & {\textbf{805}} & {\textbf{892}}\\
		\hline
		\textbf{LO31}	& \textbf{533} & {\textbf{368}} & {\textbf{100}} & {\textbf{5}} & {\textbf{801}} & {\textbf{901}}\\
		\hline
		\textbf{LO30}	& \textbf{504} & {\textbf{372}} & {\textbf{84}} & {\textbf{11}} & {\textbf{792}} & {\textbf{876}}\\
		\hline
		\textbf{HYD1}	& \textbf{594} & {\textbf{381}} & {\textbf{182}} & {\textbf{9}} & {\textbf{793}} & {\textbf{975}}\\
		\hline
		\textbf{LEW1}	& \textbf{490} & {\textbf{363}} & {\textbf{110}} & {\textbf{4}} & {\textbf{743}} & {\textbf{853}}\\
		\hline
		\textbf{SMI3}	& \textbf{475} & {\textbf{358}} & {\textbf{94}} & {\textbf{7}} & {\textbf{739}} & {\textbf{833}}\\
		\hline
		\textbf{LEW2}	& \textbf{282} & {\textbf{227}} & {\textbf{48}} & {\textbf{1}} & {\textbf{461}} & {\textbf{509}}\\
		\hline
		\textbf{TREN}	& \textbf{241} & {\textbf{182}} & {\textbf{48}} & {\textbf{2}} & {\textbf{375}} & {\textbf{423}}\\
		\hline
		\textbf{COVE}	& \textbf{202} & {\textbf{153}} & {\textbf{38}} & {\textbf{2}} & {\textbf{317}} & {\textbf{355}}\\
		\hline
		\textbf{CORN}	& \textbf{123} & {\textbf{101}} & {\textbf{19}} & {\textbf{0}} & {\textbf{205}} & {\textbf{224}}\\
		\hline
		\end{tabular}
		\end{widecenter}
		\caption{The election data and margin error bounds
			shown above are used 
                        to
			demonstrate and compare the four methods for
			calculating risk-limiting post-election audit
			sample sizes that are discussed in this paper.  The
			upper margin error bounds for the
			just-winning-losing candidate pair and the
			upper margin error bounds for \emph{all}
			candidate pairs are shown here for the Utah
			State Representative District \#~3 wide-margin
			contest in the 2004 general election. Another
			example of applying all four methods to
			determine risk-limiting sample size for a
			close-margin election contest is shown in an
			appendix.}
\end{table*}

\subsubsection*{Use the just-winning and just-losing candidate pair to calculate sample sizes}

Using accurate sampling weights, any sample size that is sufficient to detect an incorrect election outcome between the just-winning and just-losing candidate pair will have at least the same minimum probability for detecting incorrect election outcomes that may have reversed other winning-losing candidate pairs.%
\footnote{Thanks to Calandrino, Halderman, and Felten for coining the phrases ``just-winning'' and ``just-losing'' candidates.} Therefore post-election audit sample size calculations only need consider the just-winning and just-losing candidate pair error bounds and margin.

The reason that the most conservative (largest) overall post-election audit sample size is  calculated by using the win\-ning-losing candidate pair margin error bounds is because that approach produces the smallest ratio of overall vote margin to the sum of within audit unit margin error bounds
\begin{equation*}
	\frac{M}{E}
\end{equation*}

where $E_j=\sum_i{u_i}$ is the sum of all upper margin error bounds for winning-losing candidate pair $j$ over all audit units $i$ and $M = w-r$ is the overall margin. In other words, the winning-losing candidate pair with the smallest ratio $(M/E)$ is always the just-winning and justing-losing candidate pair.
	\begin{equation*}
	\frac{M}{E}=\frac{w-r}{\sum_i{b_i+w_i-r_i}}
	\end{equation*}
where $w$ is the number of total votes for the winner with the least number of votes and $r$ is the total number of votes for the losing candidate with the most number of votes (the runner-up).

A proof of this fact is provided in Appendix~E.%
\footnote{This finding is contrary to the recommendations of some
authors who recommend comparing within audit unit margin error bounds
for all winning-losing candidate pairs at once for each audit unit
when calculating sample sizes and doing discrepancy analysis \cite{Stark2008c}.}

\subsubsection*{When weighting random selections or analyzing discrepancies consider \emph{all} \\
winning-losing candidate pairs}

When weighting random selections of audit units or when analyzing discrepancies found during an audit, focus should probably not be limited to particular initial winning-losing candidate pairs.

The only type of error that can overcome the winning margin of any winning-losing candidate pair is margin error that affects that particular candidate pair. The overall total margin error found during the audit within all audit units separately for each winning-losing pair, and the maximum within audit unit margin error for all audit units separately for each winning-losing candidate pair may need consideration when analyzing discrepancies. Discrepancy error for each winning-losing pair may need to be considered, but should be considered separately for each winning-losing pair over all audit units when analyzing discrepancies found during an audit.

Table~\ref{tab:EgDataMEBs1} shows the election results data and the spreadsheet calculation of within audit unit upper margin error bounds for a specific election contest, the Utah State Representative District \#~3 wide-margin contest in the 2004 general election. Both the upper margin error bounds for the just-winning/just-losing candidate pair and for all winning-losing candidate pairs are shown.


\section{Uniform sampling methods}

Given the minimum number of miscounted audit units $C$ that could cause an incorrect outcome, then we can calculate the minimum sample size that is required to give a desired probability of having at least one such miscounted audit unit by solving the probability equation for $S$:

\begin{equation*}
 P=1-\frac{\binom{C}{0}\binom{N-C}{S}}{\binom{N}{S}}
 \end{equation*}

Therefore, the first step to determine sample sizes for risk-limiting audits that employ uniform random sampling is to determine the smallest number of miscounted audit units $C$, that could cause an incorrect initial election outcome.  Then, assuming the presence of at least $C$ miscounted audit units, the sample is sized to have at least the desired minimum probability (say 95\%+) of including at least one of these $C$ miscounted audit units.

The larger the minimum number of vote counts $C$ that it would take to cause an incorrect election, the smaller the sample size $S$ is. Thus if $C$ is overestimated, then the audit sample size may be too small to achieve the desired probability $P$ for detecting incorrect election outcomes.

\subsection*{Calculate $C$, the number of miscounted audit units that could cause an incorrect outcome}
Calculations that employ the detailed individual initial vote counts and ballots cast within each audit unit provide the most accurate sample size by providing the most accurate estimate for $C$, the minimum number of miscounted audit units that could cause an incorrect outcome.  Using the detailed data takes into account the variation upper margin error bounds that exist within different audit units.

The most accurate estimate for estimating the minimum number of miscounted initial audit units, $C$, that could cause an incorrect election outcome is obtained by ordering all initial audit units in descending order of their upper margin error bounds and then, beginning with the audit unit with the largest error bound for the just-winning and just-losing candidate pair, add a fixed proportion (the maximum level of undetectability) times their upper margin error bounds until there is sufficient error to negate the entire margin between the just-winning and the just-losing candidate pair. The value of $C$ is the number of audit units it takes to reach this level of margin error.

A program or a spreadsheet can do the calculations as follows (in pseudo-code):

\begin{itemize}
	\item 	Create an array \(\texttt{error\_bound}(i)=b_i+w_i-r_i\)  for each auditable vote count \(i = 0 \mbox{to } N-1\)
	\item Sort the \(\texttt{error\_bound}()\) array in descending order (from largest to smallest)
	\item Find the cumulative maximum possible margin error for the vote count with the largest error bound, then the two vote counts with the two largest error bounds, etc.
	For \(j=0\ to\ N-1\),\\
	 \( \texttt{CumulativeError}(j) =  \texttt{error\_bound}(0), \\ \sum_{i=0}^{1}\texttt{error\_bound}(i),\\
	  \sum_{i=0}^{2}\texttt{error\_bound}(i)\)
	\item At each step compare the cumulative maximum error to the Margin and \\
	if  \(M<k\times \texttt{Cumulative\_Error}(j)\)  \\
	then \(C \approx (j+1)\) is the minimum number of corrupt vote counts that could cause an incorrect election outcome and this value of $C$ is used to use to calculate the audit sample size, $S$.
\end{itemize}

This calculation can be performed in a spreadsheet by listing the number of ballots $b$ and vote counts for the just-winning $w$ and just-losing $r$ candidate pair within each audit unit, then calculating

\( b_i+w_i-r_i\)  for \( i = 0\ to\ N-1\) and ordering the results from largest to smallest.

This method should be used to calculate the sample size prior to making the random selections.

Once $C$ is known, we can use our desired probability $P$, and the number of total audit units $N$, we can calculate $S$ the sample size by using an accurate algebraic estimate derived by Aslam, Popa, and Rivest \cite{Aslam2007} \\
		 \(S=\left(N-(C-1)/2\right)\left(1-\left(1-P\right)^{1/C} \right) \)	  \\
or we can find $S$ exactly by using a numerical algorithm as shown by Dopp and Stenger \cite{DoppStenger}.

See Table~\ref{tab:PostElectionAuditMethodComparison} for a summary of the uniform method of calculating post-election audit sample sizes.

\subsection*{Estimate $C$ the number of miscounted audit units that could cause an incorrect outcome}

Alternatively, rough estimates for sample sizes may be obtained for planning purposes such as for estimating the funds to allocate or the number of vote count auditors that should be hired, etc.

Estimating the sample size by using the overall initial results, rather than using the initial detailed vote count (audit unit) data, in effect assumes that all vote counts contain a uniform amount of margin error. This assumption could over or under-estimate $C$ and thus over or under-estimate the sample size needed to detect an incorrect initial outcome.


\begin{table*}[htbp] 
	\label{tab:Eg1UniformMethods}
	\begin{widecenter}
	  \footnotesize
		\begin{tabular}[b]{|c|c|c|c|c|c|}
		\hline
		\multicolumn{6}{|c|}{\textbf{Example of the Improved Uniform Method \& New Uniform Estimation Method}} \\
		\hline
		& \multicolumn{2}{|c|}{\textbf{Uniform Sampling Method}} & & \multicolumn{2}{|c|}{\textbf{Uniform Estimation Method}} \\
		\hline
		{\textbf{Sample Size}} & \multicolumn{2}{|c|}{\textbf{4}} & {|c|}{} &  \multicolumn{2}{|c|}{\textbf{$5 \approx S \leq 8$}} \\
		\hline
		 {\textbf{min \#corrupt AUs}} & \multicolumn{2}{|c|}{\textbf{17}} & {} & \multicolumn{2}{|c|}{\textbf{$12 \leq C \approx 20$}}  \\
			\hline
	\textbf{Precinct}	&	 \textbf{$ku_i$} & {\textbf{cumulative margin error}}  & \multicolumn{3}{|c}{}  \\
		\cline{1-3}
	\textbf{SMI1}	& \textbf{529.6} & {\textbf{529.6}} & \multicolumn{3}{|c}{}  \\
		\cline{1-3}
		\textbf{HYD2}	& \textbf{502.0} & {\textbf{1,031.6}} & \multicolumn{3}{|c}{}  \\
		\cline{1-3}
		\textbf{SMI2}	& \textbf{501.2} & {\textbf{1,532.8}} & \multicolumn{3}{|c}{}  \\
		\cline{1-3}
		\textbf{NL04}	& \textbf{494.8} & {\textbf{2,027.6}} & \multicolumn{3}{|c}{}  \\
		\cline{1-3}
		\textbf{NL03}	& \textbf{465.2} & {\textbf{2,492}} & \multicolumn{3}{|c}{}  \\
		\cline{1-3}
		\textbf{SMI5}	& \textbf{450.8} & {\textbf{2943.6}} & \multicolumn{3}{|c}{}  \\
		\cline{1-3}
		\textbf{NLO1}	& \textbf{436.4} & {\textbf{3,380.0}} & \multicolumn{3}{|c}{}  \\
		\cline{1-3}
		\textbf{SMI4}	& \textbf{426.4} & {\textbf{3,806.4}} & \multicolumn{3}{|c}{}  \\
		\cline{1-3}
		\textbf{LO17}	& \textbf{422.0} & {\textbf{4,228.4}} & \multicolumn{3}{|c}{}  \\
		\cline{1-3}
		\textbf{NL02}	& \textbf{396.0} & {\textbf{4,624.4}} & \multicolumn{3}{|c}{}  \\
		\cline{1-3}
		\textbf{LO25}	& \textbf{378.4} & {\textbf{5,002.8}} & \multicolumn{3}{|c}{}  \\
		\cline{1-3}
		\textbf{LO04}	& \textbf{333.2} & {\textbf{5,336.0}} & \multicolumn{3}{|c}{}  \\
		\cline{1-3}
		\textbf{RCH1}	& \textbf{323.2} & {\textbf{5,659.2}} & \multicolumn{3}{|c}{}  \\
		\cline{1-3}
		\textbf{RCH2}	& \textbf{322.0} & {\textbf{5,981.2}} & \multicolumn{3}{|c}{}  \\
		\cline{1-3}
		\textbf{LO31}	& \textbf{320.4} & {\textbf{6,301.6}} & \multicolumn{3}{|c}{}  \\
		\cline{1-3}
		\textbf{LO30}	& \textbf{316.8} & {\textbf{6,618.4}} & \multicolumn{3}{|c}{}  \\
		\cline{1-3}
		\textbf{HYD1}	& \textbf{317.2} & {\textbf{6,935.6}} & \multicolumn{3}{|c}{}  \\
		\cline{1-3}
		\textbf{LEW1}	& \textbf{297.2} & {\textbf{}} & \multicolumn{3}{|c}{}  \\
		\cline{1-3}
		\textbf{SMI3}	& \textbf{295.6} & {\textbf{}} & \multicolumn{3}{|c}{}  \\
		\cline{1-3}
		\textbf{LEW2}	& \textbf{184.4} & {\textbf{}} & \multicolumn{3}{|c}{}  \\
		\cline{1-3}
		\textbf{TREN}	& \textbf{150.0} & {\textbf{}} & \multicolumn{3}{|c}{}  \\
		\cline{1-3}
		\textbf{COVE}	& \textbf{126.8} & {\textbf{}} & \multicolumn{3}{|c}{}  \\
		\cline{1-3}
		\textbf{CORN}	& \textbf{82.0} & {\textbf{}} & \multicolumn{3}{|c}{}  \\
		\cline{1-3}
		\end{tabular}
		\end{widecenter}
		\caption{This table shows the steps for using the improved and new uniform sampling methods for calculating risk-limiting post-election audit sample sizes as applied to the 2004 Utah State Representative District \#~3 wide-margin contest (Data shown in Figure~\ref{tab:EgDataMEBs1}). The confidence probability is $P = 0.99$ and the maximum level of undetectability is $k = 0.4$. The prior unimproved uniform method using the old $2sv_i$ error bounds calculated a sample size of only one (1) audit unit for this same contest because it under-est\-imates the maximum margin error that could occur within each audit unit.}
\end{table*}

Table~\ref{tab:PostElectionAuditMethodComparison} summarizes three methods and mechanisms for calculating risk-limiting election audit sample sizes.

Hence to obtain a rough estimate for the minimum number of audit units that could cause an incorrect outcome, $C$, using the overall total results, a more conservative formula for estimating both $C$ and $S$ is suggested here.

When detailed data or tools and expertise are unavailable that are needed to obtain more precise estimates, these quick estimates can be made using the overall vote totals and the number of ballots cast in the largest audit unit in the contest (this can be estimated from a prior election if it is unavailable):
             					     		       \begin{equation}
             					     		       C_{avg} = \frac{(w-r)N}{k(b+w-r)}
             					     		       \label{eq:C-avg}
             					     		       \end{equation}

  where $C_{avg}$ is the minimum number of corrupt audit units that could cause an incorrect outcome if all audit units have the average upper margin error bound of all audit units, and
             					     		       \begin{equation}
             					     		       C_{0} = \frac{b}{Nb_0}C_{avg}=\frac{b(w-r)}{kb_0(b+w-r)}
             					     		       \label{eq:C-0}
             					     		       \end{equation}
 where $C_{0}$ is the minimum number of corrupt audit units that could cause an incorrect outcome if all audit units have the maximum upper margin error bound estimated using $b_0$, the number of ballots cast in the largest audit unit in the contest, where $b$ is the number of total ballots cast in the contest and $N$ is the total number of audit units in the contest. See Appendix~A for the derivation of equation~\ref{eq:C-avg}. Equation~\ref{eq:C-0} can simply be derived by estimating the upper margin error bound of the largest precinct using the number of ballots cast,  assuming its margin is the same as the overall margin, and then solving for the ratio of $C_{avg}:C_{0}$ which always nicely reduces to $\frac{Nb_0}{b}$

 This gives us the relationship $C_{0} \leq C \approx C_{avg}$.  If $C_{0} \leq 1$ then $C = 1$. If  $C_{0} > 1$ then some further analysis may have to be done to estimate whether $C$ will be closer to the value of $C_{0}$ or $C_{avg}$.

Using a value of $k=0.40$ for estimating the sample size needed to obtain probability at least $P$ for detecting $C$ corrupt objects out of $N$ objects and then using an estimation formula for $S$ suggested by Aslam, Popa, and Rivest that sometimes slightly over-estimates the sample size needed to detect at least one miscounted audit unit if there are $C$ miscounted units,

\begin{equation}
S \approx N\left(1-(1-P)^{1/C_{avg}}\right)
\label{eq:S2-avg}
\end{equation}

\begin{equation}
S_{C_{avg}} \approx S \leq S_{C_{0}}
\label{eq:S2-0}
\end{equation}

Substituting the expression for $C_{avg}$ into the formula for $S$, and combining we get

\begin{equation}
S \approx S_{C_{avg}} = N\left(1-(1-P)^{k(b+w-r)/(N(w-r))} \right)
\label{eq:S-avg}
\end{equation}

and substituting the expression for $C_{0}$ into the formula for $S$, and combining we get

\begin{equation}
S \leq S_{C_{0}}= N\left(1-(1-P)^{kb_0(b+w-r)/(b(w-r))} \right)
\label{eq:S-0}
\end{equation}

See Appendix~A for the derivation of Equation~\ref{eq:S-avg} and Appendix~B for the derivation of Equation~\ref{eq:S2-avg}.

Other methods to estimate risk-based post-election audit sample sizes from overall initial election results can be developed using the pattern of upper margin error bounds found in prior elections' audit units. Such methods are specific to patterns found in specific election jurisdictions.


Using Formula~\ref{eq:S-avg} and~\ref{eq:S-0} to estimate
post-election audit sample sizes is simply done by using the following
steps
\begin{enumerate}
	\item Select a desired probability, $P$, for detecting the minimum level of miscount that could cause an incorrect election outcome.  (A value $P \geq 0.95$ is suggested).
\item Select an assumed maximum rate of undetectability that it is believed would not be immediately noticed if it occurs. (An initial value $k \geq 0.40$ is suggested.)
\item Estimate the minimum number of miscounted audit units, $C_{avg}$, that could cause an incorrect election outcome \\
\(C_{avg} = (N*(w-r))/((k*(b+w-r))\)
\item Calculate \(S \approx N\left(1-(1-P)^{1/C_{avg}}\right) \) using a spreadsheet or a calculator.
\end{enumerate}

It only takes one row and five columns in a spreadsheet to estimate the risk-limiting post-election audit sample size for various election contests using this new formula.%
\footnote{A spreadsheet formula for calculating $C$, the number of corrupt counts that could cause an incorrect election outcome is \\
 \texttt{``=(N*(w-r))/(k*(b+w-r))''}.
 A spreadsheet formula for calculating $S$, the audit sample size needed to provide $P$, probability for detecting one or more corrupt vote counts if there are C corrupt vote counts is  \\
\texttt{``=N*(1-(1-P)\char`\^(1/C))''}.}

\section{Weighted sampling methods}

When random selections are weighted by margin error bounds, the
probability proportional to margin error bound (PPMEB) method can be
used to determine the risk-limiting election audit sample size
\cite{Aslam2008,DoppStraight,Dopp2007-8b,Calandrino2007a,Dopp2007-8c}. 

Probability proportional to margin error bound methods for post-election auditing reduce the post-election audit sample size necessary to achieve a desired detection probability by weighting the selections of audit units using the audit unit margin error bounds \cite[p.10 and 16]{Aslam2008}, \cite{Dopp2007-8b,Dopp2007-8c,Calandrino2007a}.

This article proposes new weights and probabilities for randomly selecting audit units that will work well for both individual ballot audit units proposed by Calandrino, Halderman, and Felten and for variable sized audit units proposed by Aslam, Popa, and Rivest \cite{Aslam2008,Calandrino2007a}.

In calculating the random sampling weights and probabilities, it is probably best to avoid making assumptions about which initial winning-losing candidate pair could be incorrect. Thus the upper margin error bounds that each ballot could contribute to margin error for \emph{any} winning-losing candidate pairs should be considered developing sampling weights.

To ensure the most conservative approach (to ensure an adequate sample size), the margin in total ballots to be overcome is still the margin between the initial just-winning and the just-losing candidate pair.

This method applies equally well to any size audit unit and to election contests with any number of seats being elected. The sample size is calculated from the probabilities that each audit unit is randomly selected.

Appendix~D describes some flaws with the upper margin error bounds and the sampling weights proposed in some other authors' recommendations.

\subsection*{Calculate the random selection weights}

Note that each initial losing candidate $j$ would certainly prefer that the sampling weights used to select audit units are the losing candidate's own upper margin error bounds with an initial winning candidate as follows:
\begin{equation*}
u_i=b_i+w_{ij}-l_{ij}
\end{equation*}
because then initial votes cast for that particular losing candidate would tend to escape scrutiny, while votes of other candidates would receive more scrutiny.

However, in order to provide equal treatment to all losing candidates when weighting random selections, we avoid using upper margin error bounds for particular winning-losing candidate pairs and thus avoid any assumption about which initial winning-losing candidate pair(s) are incorrectly reported.

If a perpetrator with good insider access knows that the just-winning and just-losing upper margin error bounds are used to weight selections, that could provide a strategy to hide miscount by repositioning the relative order of candidate vote totals. In other words, a perpetrator could try to cause the initial results to show a different ordering of the candidates to reduce the chances of scrutiny of certain candidates' votes.

Hence we need a within audit unit upper margin error bound for \emph{all} winning-losing candidate pairs for weighting random samples.

Not that there may be exceptions to this rule when auditors are increasing an audit sample in response to detecting errors in certain candidates' votes and believe that certain candidates' votes need the most scrutiny.

The selection weights for audit units suggested here allow for the possibility that any initial losing candidate might be a rightful winner, and any initial winning candidate could be a rightful losing candidate by using the upper margin error bounds for each audit unit $i$ of:

\begin{equation}
u_i= 2\sum_i{w_i}+\sum_i{l_i}
\label{eq:weights}
\end{equation}
where
\begin{align*}
l_i &= \text{the number of votes for any losing candidates} \\
 & \text{\& under-over-votes,}\\
w_i &= \text{the number of votes for any winning candidate,} \\
w-r &= \text{the overall initial margin of just-winning and}\\
 & \text{just-losing candidate pair.}
\end{align*}

See Appendix~F for an example showing how this maximum amount of margin error could occur within an audit unit.

Another benefit of this method is that if there is more than one winner in a multi-seat election, the sampling weights suggested here do not assume which winner may be an incorrect winner, and do not assume which initial losing candidate may be a rightful winning candidate.

This weighted sampling proposal is consistent with the amount of upper margin error that each ballot can contribute to causing an incorrect outcome, given that we do not known which initial winning-losing candidate pair may be incorrect before auditing.

\subsection{The Improved PPMEB Method}

Probability proportional to margin error bound (PPMEB) methods use margin error bounds for sampling weights.

One PPMEB method for sampling individual ballots looks at how many of each particular type of audit unit with its own particular voting patterns could be used to cause an incorrect election outcome \cite{Calandrino2007a}.  This section suggests using new more accurate sampling weights for the approach first developed by Calandrino et al.\ and generalizes the method to any size audit unit.%
\footnote{Appendix~D explains why the original method proposed by Calandrino et al.\ for random selection weights do not work as well as the improved method presented in this paper.}

Using this improved method, the fewest number of such miscounted audit units that could possibly cause an incorrect election outcome for each type of audit unit $1\leq i \leq n$ is:
\begin{equation}
c_i\approx \frac{(w-r)}{k(2\sum_i{w}+\sum_i{l})}
\label{eq:corrupt-units}
\end{equation}

$c_i$ is an estimate for the number of similar audit units it would take to alter the election outcome by taking the maximum reasonable proportion of the over margin between the just-winning and just-losing candidate pair that could be eaten up by miscounted ballots with this vote pattern without raising immediate suspicion.

Therefore, the probability that each audit unit (individual ballot) \\
$1\leq i \leq n$ should be selected for auditing is:

\begin{equation}
p_i=1-(1-P)^{1/c_i} \text{and substituting for $1/c_i$ we get}
\label{eq:ProbEq}
\end{equation}

\begin{equation}
p_i=1-(1-P)^{k(2\sum_i{w_i}+\sum_i{l_i})/((w-r))}
\label{eq:ProbEq3}
\end{equation}
where within each audit unit,
 \begin{align*}
l_i &= \text{the number of votes for losing candidates} \\
 & \text{~\& under \& over-votes,}\\
w_i &= \text{the number of votes for any winning candidate,} \\
w-r &= \text{the overall initial margin of just-winning,}\\
 & \text{just-losing candidate pair, and}\\
 k &= \text{the assumed maximum level of undetectability.}
\end{align*}

\subsubsection*{The sample size}
The expected value for the overall post-election audit sample size is equal to the sum of the probabilities that each audit unit is selected for auditing:
\begin{equation}
\sum_i{p_i}=\sum_i{(1-(1-P)^{k(2\sum_i{w_i}+\sum_i{l_i})/(w-r)})}
\label{eq:ProbEq2}
\end{equation}


\subsection{The Improved PPMEBWR Method}

Another PPMEB approach looks at how much each audit unit could contribute overall to causing an incorrect election outcome. The probability proportional to margin error bound with replacement (PPMEBWR) approach was first developed by Aslam, Popa, and Rivest.\cite[p. 16]{Aslam2008}

This section improves upon the method recommended by Aslam et al.\ by using the more precise just-winning/just-losing candidate pair upper margin error bounds to calculate the number of random draws and by using the new all-candidate-pair sampling weights presented here to calculate the sampling weights for the PPMEBWR method.%
\footnote{
Appendix~D explains why the original Aslam et al.\ upper margin error bounds and random selection weights do not obtain the stated probability for detecting incorrect election outcomes.}

Overall the PPMEBWR method is simply described as follows:

The number of random draws $t$ with replacement of audit units is
\begin{equation*}
t=\ln(1-P)/\ln(1-M/E_{wr})
\end{equation*}

where $M$ is the overall vote margin for the just-winning and just-losing candidate pair, and $E_{wr}=k\sum{u_i}$, the maximum level of undetectability $k$ times the sum of the within audit unit upper margin error bounds for the just-winning and just-losing candidate pair. In other words, $E_{wr}=k\sum_i{(b_i+w_i-r_i)}$ where $b_i$ is the total number of ballots cast, $w_i$ is the number of initial votes counted for the just-winning candidate, and $r_i$ is the number of initial votes reported for the just-losing candidate in audit unit $i$, and where $P$ is the desired probability that there is at least one miscounted audit unit in the sample if the initial reported election outcome is incorrect. $E_{wr}$ simply reduces to $E_{wr} = k(b + w - r)$ or $k$ times the total number of ballots + total votes for the just-winning candidate minus the total votes for the just-losing candidate in the contest because $k\sum_i{(b_i+w_i-r_i)} = k(\sum{b_i} + \sum{w_i} - \sum{r_i})$.

So, as not to bias the sample in favor of a particular initial reported losing candidate over another, the probability $p_i$ for sampling each audit unit $i=1$ to $i=N$ is
\begin{align*}
p_i &= \frac{2\sum_{ij}{w_{ij}}+\sum_{ij}{l_{ij}}}{E_a}  \\
p_i &= \frac{b_i + \sum_{ij}{w_{ij}}}{E_a}
\end{align*}
where the sum of margin error bounds is $E_a=\sum_i{(2\sum_{ij}{w_{ij}}+\sum_{ij}{l_{ij}})}$ the sum of the upper margin error bounds for \emph{all} winning-losing candidate pairs, and where the number of votes for winning candidates $j$ in audit unit~$i$ is $w_{ij}$ and the number of votes for losing candidates $j$ and under-votes in audit unit $i$ is $l_{ij}$.

\subsubsection*{First step: Sum the total error bounds}

The first step is to calculate $E_{wr}=k\sum_i{(b_i+w_i-r_i)}$ the sum of the error bounds for the $i=1$ to $i=N$ audit units for the just-winning and just-losing candidate pair. We multiply the upper error bounds times $k$, the maximum level of undetectability where $0\leq k \leq 1$ as discussed above because not all ballots can be miscounted or it would be immediately evident without an audit.  This sum can be easily calculated by adding the total number of ballots cast in the election contest, plus the total number of initial votes counted for the just-winning candidate, minus the total number of initial votes counted for the just-losing candidate.

Also calculate $E_a=\sum_i{(2\sum_{ij}{w_{ij}}+\sum_{ij}{l_{ij}})}$ the sum of the upper margin error bounds for \emph{all} winning-losing candidate pairs for the $i=1$ to $i=N$ audit units by summing two times the total initial votes counted for any winning candidate plus the total initial votes counted for any losing candidate.

\subsubsection*{Second step: Calculate the number of draws}

Calculate $t$ using the margin in ballots between the just-winning and just-losing candidate pair, $M=w-r$, and $E$, and $P$ the desired confidence probability that there will be at least one miscounted audit unit in our sample if the initial election outcome is incorrect.
\begin{equation}
t=\ln(1-P)/\ln(1-M/E_{wr})
\end{equation}
The derivation for the number of draws for the weighted sampling method is described in previous literature.\cite[p. 10]{Aslam2008}, \cite{Dopp2007-8b}.

\subsubsection*{Third step: Calculate the selection weights}

Now calculate the sampling weights for each of the $i=1$ to $i=N$ audit units.

\begin{equation*}
p_i=\frac{2\sum_{ij}{w_{ij}}+\sum_{ij}{l_{ij}}}{E_a}  \texttt{or} \\
p_i=\frac{b_i + \sum_{ij}{w_{ij}}}{E_a}  \texttt{or}
\end{equation*}

using the upper margin error bounds for \emph{all} candidates in audit unit $i$ as shown in equation~\ref{eq:weights}.

The expected value for the sample size $S$ will be:
\begin{equation*}
S=\sum_{i}(1-(1-p_i)^t)
\end{equation*}

\begin{table*}[htbp] 
	\label{tab:EgCompareWeightedMethods1}
	\begin{widecenter}
	  \footnotesize
		\begin{tabular}[b]{|c|c|c|c|c|c|}
		\hline
		\multicolumn{6}{|c|}{\textbf{Example applying the Improved Weighted Sampling Methods}} \\
		\hline
		& \multicolumn{2}{|c|}{\textbf{Improved PPMEB approach}} & & \multicolumn{2}{|c|}{\textbf{Improved PPMEBWR approach}} \\
		\hline
		{\textbf{Expected Sample Size}} & \multicolumn{2}{|c|}{\textbf{6}} & & \multicolumn{2}{|c|}{\textbf{3}} \\
		\hline
		 {\textbf{}} & {\textbf{}} & {\textbf{}} & {} & {\textbf{\#Draws}} & {\textbf{3}}  \\
		 \hline
	\textbf{Precinct}	&	 \textbf{$c_i$} & {\textbf{$p_i$}}  & {} & {\textbf{$p_i$}} & {\textbf{$1-(1-p)^t$}}  \\
		\hline
	\textbf{SMI1}	& \textbf{11.07} & {\textbf{0.34}} & {} & {\textbf{0.07}} & {\textbf{0.18}} \\
		\hline
		\textbf{HYD2}	& \textbf{10.48} & {\textbf{0.36}} & {} & {\textbf{0.07}} & {\textbf{0.19}} \\
		\hline
		\textbf{SMI2}	& \textbf{11.78} & {\textbf{0.32}} & {} & {\textbf{0.06}} & {\textbf{0.17}} \\
		\hline
		\textbf{NL04}	& \textbf{11.45} & {\textbf{0.33}} & {} & {\textbf{0.06}} & {\textbf{0.18}} \\
		\hline
		\textbf{NL03}	& \textbf{12.79} & {\textbf{0.30}} & {} & {\textbf{0.06}} & {\textbf{0.16}} \\
		\hline
		\textbf{SMI5}	& \textbf{13.06} & {\textbf{0.30}} & {} & {\textbf{0.06}} & {\textbf{0.16}} \\
		\hline
		\textbf{NLO1}	& \textbf{12.86} & {\textbf{0.30}} & {} & {\textbf{0.06}} & {\textbf{0.16}} \\
		\hline
		\textbf{SMI4}	& \textbf{13.81} & {\textbf{0.28}} & {} & {\textbf{0.05}} & {\textbf{0.15}} \\
		\hline
		\textbf{LO17}	& \textbf{14.08} & {\textbf{0.28}} & {} & {\textbf{0.05}} & {\textbf{0.15}} \\
		\hline
		\textbf{NL02}	& \textbf{15.22} & {\textbf{0.26}} & {} & {\textbf{0.05}} & {\textbf{0.14}} \\
		\hline
		\textbf{LO25}	& \textbf{16.08} & {\textbf{0.25}} & {} & {\textbf{0.04}} & {\textbf{0.13}} \\
		\hline
		\textbf{LO04}	& \textbf{17.21} & {\textbf{0.23}} & {} & {\textbf{0.04}} & {\textbf{0.12}} \\
		\hline
		\textbf{RCH1}	& \textbf{18.63} & {\textbf{0.22}} & {} & {\textbf{0.04}} & {\textbf{0.11}} \\
		\hline
		\textbf{RCH2}	& \textbf{18.73} & {\textbf{0.22}} & {} & {\textbf{0.04}} & {\textbf{0.11}} \\
		\hline
		\textbf{LO31}	& \textbf{18.55} & {\textbf{0.22}} & {} & {\textbf{0.04}} & {\textbf{0.11}} \\
		\hline
		\textbf{LO30}	& \textbf{19.08} & {\textbf{0.21}} & {} & {\textbf{0.04}} & {\textbf{0.11}} \\
		\hline
		\textbf{HYD1}	& \textbf{17.14} & {\textbf{0.24}} & {} & {\textbf{0.04}} & {\textbf{0.12}} \\
		\hline
		\textbf{LEW1}	& \textbf{19.59} & {\textbf{0.21}} & {} & {\textbf{0.04}} & {\textbf{0.11}} \\
		\hline
		\textbf{SMI3}	& \textbf{20.06} & {\textbf{0.21}} & {} & {\textbf{0.04}} & {\textbf{0.10}} \\
		\hline
		\textbf{LEW2}	& \textbf{32.83} & {\textbf{0.13}} & {} & {\textbf{0.02}} & {\textbf{0.06}} \\
		\hline
		\textbf{TREN}	& \textbf{39.50} & {\textbf{0.11}} & {} & {\textbf{0.02}} & {\textbf{0.05}} \\
		\hline
		\textbf{COVE}	& \textbf{47.07} & {\textbf{0.09}} & {} & {\textbf{0.01}} & {\textbf{0.05}} \\
		\hline
		\textbf{CORN}	& \textbf{74.6} & {\textbf{0.06}} & {} & {\textbf{0.0}} & {\textbf{0.03}} \\
		\hline
		\end{tabular}
		\end{widecenter}
		\caption{This table shows the steps for using the improved weighted sampling methods for risk-limiting post-election audits applied to the 2004 Utah State Representative District \#~3 wide-margin contest (data shown in Table~\ref{tab:EgDataMEBs1}). This example uses a confidence probability of $P=0.99$ and a maximum level of undetectability of $k=0.4$. In this case, the improved PPMEBWR method shows an expected sample size of 3 audit units versus the old PPMEBR method that used the $2sv$ error bound that would have calculated a sample size of only 1 audit unit. The improved PPMEB method is more conservative than the improved PPMEBWR method with an expected sample size of 6 audit units.}
\end{table*}

\begin{table*}[htbp] 
	\caption{Summary of Improved Methods for Risk-Reducing Post-Election Audit Sampling}
	\label{tab:PostElectionAuditMethodComparison}
	\begin{widecenter}
	\footnotesize
		\begin{tabular}[b]{|p{4cm}|p{6cm}|p{5cm}|}
		\hline
		& \multicolumn{2}{c}{\textbf{Random Selection Method}} \\
		\cline{2-3}
		\textbf{Data \& Tools Available?}
		&\textbf{Uniform Probability Distribution}
		& \textbf{Probability Proportional to Margin Error Bound (PPMEB)} \\
		\hline
		\textbf{Detailed initial audit unit (vote count) data and a spreadsheet or computer program are available}
		& {\raggedright For each vote count $i$ \\
		\( \texttt{error\_bound}_{wr}(i)=k(w_{i}-r_{i}+b_{i})\)\\
		where \(0.4\leq k \leq 1 \) for the just-winning/just-losing candidate pair. Order \( \texttt{error\_bound} \) in descending size order. Then for each $j$ from $0$ to $N-1$, if \(w-r<\sum_{i=0}^{j} \texttt{error\_bound}(i)\) then stop and \(C=j+i\). Then the sample size $S$ can be found for detecting $C$ corrupt audit units using a precise numerical method \cite{DoppStenger} or by using an estimate \cite{Aslam2007} \\
		 \(S=\left(N-(C-1)/2\right)\left(1-\left(1-P\right)^{1/C} \right) \)}
 &
 {\raggedright \emph{PPMEB METHOD} \\If miscounted, each audit unit $i$ with a particular vote pattern would take approximately at most $c_i\approx \frac{(w-r)}{k(b_i+\sum_i{w})}$ of such units to overcome the closest contest margin. So select each audit unit with probability \\
 \(p_{i}=1-(1-P)^{1/c_i}\) \\
 The expected value for the sample size is \\
  $S= \sum_i{p_i}$ \\
  \emph{PPMEBWR METHOD} \\
   \(E_{wr}= k(b + w - r)\) the sum of k times the upper margin error bounds for the just-winning/just-losing candidate pair. \\
 For each audit unit $i$ \( \texttt{error\_bound}_{a}(i)=b_i + \sum_{ij}{w_{ij}}\)
 where \(E_{a}=\sum_{i=0}^{N-1} \texttt{error\_bound}_{a}(i) \) is the sum of bounds for \emph{all} winning-losing candidate pairs.
 The selection probability for each audit unit is \\
 $p_i=\frac{\texttt{error\_bound}_{a}(i)}{E_a}$ The number of selection rounds is \\
 \(t=\ln(1-P)/\ln(1-M/E_{wr})\) and the expected sample size is \\
  \(S=\sum_{i}(1-(1-p_i)^t) \)}
 \\
 \hline
		\textbf{Overall, but not detailed, election results data are available}
		&	{\raggedright Estimate the number of corrupt vote counts to cause an incorrect election outcome\\
		 \(C_{avg} = ((w-r)N)/(k(b+w-r))\) \\
		 $ C_0 = \frac{b}{Nb_0}C_{avg}$ \\
		 then the sample size is between these two values \\ \( N\left(1-(1-P)^{1/C_{avg}}\right) \leq S \leq  N\left(1-(1-P)^{1/C_{0}}\right) \)\\
		  where \(0.5\leq k \leq 1\). \\
		  See Appendices A \& B for more information.}
		& Not possible \\ \hline
		\end{tabular}
		\end{widecenter}
\end{table*}

\section{Other Sample Size Considerations}
\label{sec:OtherConsiderations}
\subsection*{Losing candidates select additional audit units}
To achieve the desired minimum probability for detecting incorrect
election outcomes, due to the use of the assumption of a maximum level
of undetectability, $k$, we must allow losing candidates or their
representatives to select at least one additional discretionary audit
unit to supplement the randomly-selected sample. Discretionary audit
units are necessary due to basing random sample sizes on the
assumption that suspicious-looking audit units with more than a fixed
level of margin error $k$, say 40\% or 50\%,
would be investigated without the necessity for an audit
\cite[p. 14]{DoppStenger}, \cite{DoppStraight},
\cite{Dopp2007-8a}, \cite[p. 7]{Stark2008b},
\cite{Appel,Hall2008a}.


This crucial practice would be thwarted if the losing candidates
  were required to pay the costs of   any discretionary audits and if these discretionary audits were  separately administered. Risk-limiting election audits that do not allow for the selection of  discretionary audit units  over-state the confidence probability that the audit will detect  incorrect outcomes because  they in essence put no upper limit on the margin error that could  occur within audit units,  negating the assumptions of their own  sample size calculations.

Discretionary audit units should be included as part of the initial manual audit without cost to candidates, or a risk-limiting audit may fail to achieve its stated minimum probability to detect erroneous initial outcomes.

\subsection*{Select one additional audit unit from each ``missed'' jurisdiction}
Unless at least one audit unit is sampled from each separately administered election jurisdiction where an election contest occurs, innocent ballot programming errors, voting system problems, or fraud that is peculiar to one jurisdiction could be missed. It is important to make these additional random selections only \emph{after} the initial random selections are made from any missed jurisdictions because otherwise audits would insufficiently sample high-population areas, thus providing a map for potential perpetrators for what areas to target in order to increase the chance that audits would not detect the miscount.%
\footnote{Attorney Paul Lehto pointed this out in emails.}

\subsection*{Size audit units as uniformly as possible}
It is important to keep the size variation of auditable vote counts as small as possible for two reasons:
\begin{enumerate}
	\item	Wide variation in the number of ballots cast (and the margin error available) in different audit units can result in sufficient margin error to cause an incorrect election outcome existing in just a few of the largest audit units. Especially if uniform sampling methods are used (as unfortunately is required by some state's auditing statutes) outcome-altering error may be missed and wide variation in audit unit sizes increases the need for manually counting more ballots.
\item	Risk-limiting audits can be conducted more efficiently if the total number of ballots in all audit units is roughly uniform by evening out within audit unit margin error potential somewhat. To achieve the same risk level, it requires manually counting fewer audit units overall when the audit unit sizes are more uniform.%
\footnote{With large size variation of vote counts the Uniform method for estimating sample sizes without detailed data will underestimate the sample size needed to achieve the desired probability of detecting incorrect outcomes, and thus under-states the maximum risk level for a calculated sample size.}
\end{enumerate}

\subsection*{Small-sized audit units are more efficient}
The smaller the number of total ballots in the reported audit units, the fewer the number of ballots overall will need to be manually audited to provide the desired probability for detecting incorrect outcomes (\cite{Wand,Walmsley,Atkeson}).

Voting system design and procedural obstacles to conducting effective, efficient audits because today almost all commercial voting system tabulators are designed to only report precinct vote counts and do not report which machines tallied those votes and do not report tallies for ballots that are counted and stored together. These design flaws make precincts or polling places the only audit unit that election officials can conveniently use for auditing election results accuracy without taking time-consum\-ing extra measures that delay the tallying and public release of initial election results just at the time when candidates and press are anxiously waiting for results \cite{Dopp2009c}.

Election officials may wait to begin a post-election audit after all provisional and mail-in ballots have been counted and publicly reported, or alternatively officials could use somewhat inconvenient, time-consuming ways to count and to publicly report mail-in and provisional ballots in batches that are roughly equal in size to the median or average-sized audit units. See the upcoming article in this same series \citetitle{Checking Election Outcome Accuracy --- Post-Election Audit Procedures}.

\subsection*{Some errors to avoid}

Some authors state that risk-limiting post-election audits can be
performed using any initial sample size
\cite[p. 18]{Stark2008a}, \cite{Stark2009b,Hall2009}. However, audits
that use insufficient initial sample sizes are either ineffective or
inefficient --- Ineffective because insufficient initial samples are
not likely to detect well-hidden vote fraud in cases when a minimum
number of audit units are miscounted to cause an incorrect
outcome and yet avoid a state-mandated recount; and inefficient and administratively burdensome because even
if no discrepancies are found in a too-small sample,
limiting the risk requires manually auditing another round of randomly selected audit units.

Because keeping $w-r$ constant and decreasing $E$ increases the quantity $M/E$ and thus decreases the post-election audit sample sizes, procedures that:
\begin{enumerate}
	\item take minimums of within audit unit upper margin error bounds out of all the winning-losing pairs produce an insufficient sample size,
	\item take minimums of values that are less than the margin error bounds for the just-winning and just-losing candidate pair when calculating the total upper margin error bound $E$, produce an insufficient sample size,
	\item use proportions of total votes or of total ballots to approximate the total upper margin error bound $E$ (such as $\sum{2s_iv_i}$) that are more often less but sometimes more than the actual margin error bounds produce insufficient sample sizes and poor sampling weights. (See Appendix~D for a discussion.)
\end{enumerate}

Methods that produce insufficient sample sizes will not achieve their stated minimum probability for detecting vote fraud that occurs by miscounting a minimum number of audit units to cause an incorrect outcome, unless the audit sample is expanded even when no discrepancies are found in the initial sample.

Failure to use and to understand the logical implications of using a maximum level of undetectability of less than one (1) in any of the methods described in this paper can cause:
\begin{itemize}
	\item a test result that says to expand an audit sample unnecessarily, in some cases even after a 100\% manual count has already been performed; and
 \item an unmerited expansion of the sample size when there is only a one ballot discrepancy.~\cite[p. 6--9]{Stark2009a}
 \end{itemize}


When calculating sample sizes, sampling weights, or analyzing discrepancies, methods that use a different winning-losing candidate pair's margin error bound for each different audit unit or even mix up the data for different election contests into one calculation method \cite{Stark2008a,Stark2008b,Stark2009b,Hall2009} are less precise. Such methodology does not save significant computation time or resources because the data and resources used to do more precise calculations are about the same.  The sum or maximum of within audit unit upper margin error bounds for each winning-losing candidate pair separately could just as easily and more accurately be calculated for one election contest at a time, and then that sum or maximum compared with those of other winning-losing candidate pairs for each separate election contest. The method of mixing up multiple election contests into one calculation will initially over-audit some contests and under-audit others, making the discrepancy analysis conclusions less efficient for some contests and less precise, possibly causing unnecessary false positives or failures to detect some incorrect election outcomes. Also trying to manually count multiple contests on the same ballot at the same time during an audit, is likely to negate the efficiencies of using the sort and stack method for counting paper ballots \cite{Stevens2007a}.

All election contests require a sample size larger than zero (0). In fact the formula given to show that in some cases a post-election audit is not required to confirm an outcome \cite[p. 10]{Stark2008a}, \cite[p. 6]{Stark2009a} can be instead be used to prove that in any contest with more than one candidate, the sample size must be greater than zero to confirm the outcome because winning candidates have more initial reported votes than losing candidates.

\section{Summary \& Recommendations}

Risk-limiting post-election audits limit the risk that an incorrect election outcome, the wrong winner, is incorrectly certified to any desired small maximum probability.%
\footnote{Risk-limiting audits should limit risk of incorrect outcomes even in cases where there are a large number of under-votes. In the 2006 Sarasota County, Florida Congressional District 13 race, there were
 18,000 missing votes (undervotes) in a Democratic-leaning county recorded on paperless ES\&S DREs in a tight election. Statistics show that these undervotes probably altered the outcome, causing Christine Jennings to lose the US House race.} Table~\ref{tab:PostElectionAuditMethodComparison} summarizes the improved methods presented in this paper for calculating post-election audit sample sizes and for weighting random selections that ensure that post-election audits achieve the desired minimum probability of detecting and correcting any incorrect initial outcomes. 
 
The new upper margin error bounds for all winning-losing candidate pairs that are presented in this article improve the sample size calculations and sampling weights of existing approaches and work for any size (1-ballot or many-ballot) audit units, for single or multi-winner election contests and for approaches treating all losing candidates equally. Using precisely correct upper margin error bounds ensures the adequacy of post-election audit sample sizes and allows random selection weights focusing on all winning-losing candidate pairs or on a particular winning-losing candidate pair.

When weighting random selections, this author recommends using the more conservative PPMEB method that uses a larger post-election audit sample size. Both the PPMEB and the PPMEBWR weighted sampling methods can be generalized for any sized audit units (from 1 ballot to many ballots) using the improved methods suggested in this paper.

Methods and materials for auditing elections and training auditors still need development. Election officials, vote count auditors, election integrity advocates, and voting system vendors, would benefit from
\begin{itemize}
	\item	Better voting system design specifications and technical features would make voting machines and tabulators more audit-able and accountable, provide more convenient methods to check vote count accuracy and to determine how, when, and where errors occur and the cause of errors.
	\item 	An accurate, understandable post-election auditing manual and an easy-to-use tool-kit that includes a clear explanation of and a program for calculating risk-limiting election audit sample sizes, including procedures for conducting effective and efficient post-election audits. This manual should provide pictures, forms and toolkits; and explain election auditing in simple-to-follow terms for lay persons%
	\footnote{In order to be successful, so that election jurisdictions do not have to hire statisticians to plan every post-election risk-limiting audit, such a project would require professional manual writers, and open source computer program developers, to create an easy-to-use manual and tool-kit in collaboration with election officials, security experts, and election auditing experts.}.
	\item Use of open publicly defined computer data recording format standards uniformly adopted by all election districts to provide consistent access to all electronic ballot records and making voting system components, including auditing devices, inter-operable%
	\footnote{To date only the Secretary of State office in California has reported precinct level election results using international recording standards.}.
	\item 	Conferences that bring together experts in election auditing methods together with State and county election officials and State and Federal legislators.
	\item 	Clear, easy-to-follow instructions and computer programs for making verifiable fair, weighted or uniform random selections of audit units.
	\item   Methods to generate audit vote fraud and discrepancy test data to test the ability of various audit methods' to detect various vote fraud strategies.
	\item 	A complete set of methods, tools, and decision-mak\-ing algorithms for analyzing post-election audit discrepancies, including algorithms for deciding whether to certify an election outcome or to expand the audit sample.%
\footnote{Most current discrepancy analysis methods expand
          post-election audit sample sizes whenever audit discrepancy exceeds an
          arbitrary fixed amount or rate of error and otherwise certify the
          election outcome. These approaches could incorrectly certify very
          close margin contests or unnecessarily expand audits in wide-margin
          contests. The third article in this series \citetitle{Checking
          Election Outcome Accuracy --- Post-Election Audit Discrepancy
          Analysis} covers algorithms and methods for analyzing post-election
          audit discrepancies.}
	\item The development of precise methods to assist losing candidates in selecting discretionary ``suspicious-looking'' audit units to add to the randomly selected sample.
	\item A college textbook or textbook chapter to explain post-election auditing methodology and principles.
\end{itemize}

The knowledge, resources and skills needed to implement routine post-election risk-limiting audits that provide high confidence in the accuracy of final election outcomes need on-going development and dissemination.

There are two other articles in this series \citetitle{Checking Election
Outcome Accuracy --- Post-Election Audit Discrepancy Analysis}, and
\citetitle{Checking Election Outcome Accuracy --- Post-Election Audit
Procedures}.

\section*{Acknowledgments}
Thanks to David Webber of Open Voting Solutions%
\footnote{http://openvotingsolutions.com}
for reviewing this article and making helpful suggestions and to
Political Science Professor Andrew Gellman for suggesting the addition
of an introduction describing the work of other authors in the field
of post-election auditing methods. Thanks to the contributions of
other authors who formerly developed weighted sampling approaches for
calculating risk-limiting post-election audit sample sizes or who
derived formulas for estimating uniform sample sizes that this paper
advances and improves upon.  Profuse thanks to Nelson Beebe for taking many patient hours teaching me how to use LaTeX and to improve this article's formatting. Thanks to Mathematics and
Computer Science Professor Frank Stenger for making a crucial
stylistic suggestion and to Mathematics Professor Stewart Ethier for
valuable formatting and LaTeX suggestions, and to other dedicated
election integrity advocates who motivated me to keep working on and
improving this paper.

\onecolumn
\section*{Appendix~A: Derivation of Estimate for the Minimum Number of Corrupt Vote Counts to Alter an Outcome}

The smallest amount of margin error that could change the reported outcome is the actual margin in ballots between the winning candidate with the smallest number of votes (the just-winning candidate) and the losing candidate with the most votes (the just-losing candidate).

\subsection*{The upper bound for margin error}

The upper bound in number of ballots for the total margin error that
could contribute to an incorrect election outcome is given by the
expression \(b+w-r\), or as a percentage of ballots, is
\(1+\frac{w-r}{b} \), where $b$ is the total number of ballots cast in
the election contest, $w$ is the reported number of votes counted for
the winner, and $r$ is the number of reported votes counted for the
closest runner-up.

\subsection*{Estimating the minimum number of corrupt vote counts, $C$ that could cause an incorrect outcome}
\subsubsection*{$C_avg$, the estimate if margin error bounds were all average sized}

One way to estimate the minimum number of corrupt audit units, $C$, that could cause an incorrect winner, is to divide the margin in ballots between the just-winning and just-losing candidates by the average upper margin error bound for all the reported vote counts. This gives a measure for how many corrupt audit units could cause an incorrect election outcome if all audit units have the average upper margin error bounds.

\begin{equation*}  \frac{TotalMarginError2ChangeOutcome}{AverageMarginErrorPerAuditUnit}=\#AuditUnits2ChangeOutcome
\end{equation*}

This estimates the number of audit units with sufficient possible margin error to cause an incorrect election outcome.

Thus, this estimate for $C_{avg}$ is

\begin{equation*}
 C_{avg}\approx \frac{w-r}{\frac{k(b+w-r)}{N}}
  \end{equation*}
  which reduces to
 \begin{equation*}
  C_{avg}\approx \frac{N(w-r)}{k(b+w-r)}
   \end{equation*}
    However, this method usually underestimates $C$ in real elections because the average amount of margin error in all audit units is always less than the amount of possible margin error in the largest audit units.  Hence using a larger $k$ value such as \(k\geq 0.50\) will help to compensate by finding a more conservative estimate (larger sample size) for $S$.

    \subsubsection*{$C_0$, the most (overly) conservative estimate}

    From the above formula for $C_{avg}$and some estimates for the reported margin in the largest audit unit, a formula can be derived and simplified that provides a more conservative (smaller) estimate for the number of corrupt audit units $C_0$ and thus provides a more conservative (larger) sample size estimate.

    Simply use the relationship that
    \begin{displaymath}
     C_{0} \approx \frac{bC_{avg}}{Nb_0}
             \qquad \parbox{0.5\textwidth}{\raggedright where $b_0$ is
             the number of ballots in the largest audit unit with the
             most ballots cast}
    \end{displaymath}

   This estimate assumes that the margin in the largest audit unit is
   the same as the overall margin in the election contest.  That is,
   the margin error bound in the largest audit unit is
   $b_0(1+(w-r)/b)$ so that the ratio of the margin error in the
   largest audit unit to the audit unit with average margin error
   algebraically reduces to
   $Nb_0/b$.

\section*{Appendix~B: Derivation of a Uniform Election Audit Sample Size Estimate}
This derivation has been previously described elsewhere \cite{Rivest2007,Dopp2007a}. Two well-known mathematical facts are used in the derivation:
\begin{enumerate}
	\item For values of $0 < c < 1$, \( (1-c)\approx e^{-c}\) so
              that \( (1-c)^{S}\approx e^{-cS}\) and therefore taking the natural
              log of both sides \( \ln(1-c)^{S}\approx -cS\);
        \item
              The formula for estimating the number of distinct
              elements, $S$, in a sample of size $t$ drawn (with
              replacement) from a set of size $N$ is \(S\approx
              N(1-e)^{-t/N}\).
\end{enumerate}

We begin with $N$ total vote counts, of which $C$ are corrupt (miscounted) and ask what randomly selected sample size S will give us at least probability $P$ for having one or more corrupt vote counts. Estimating the desired probability from sampling with replacement (an easier equation to solve than sampling without replacement), we find the probability of not detecting any miscount. Because, $P(0)+P(1)+\cdots+P(S) = 1$ and $S \leq C$, \\
then the probability of drawing one or more corrupt vote counts is
\begin{equation*}
P(1)+P(2)+\cdots+P(S) = 1- P(0).
\end{equation*}

If we randomly draw one vote count, then the chance of drawing a
corrupt vote count is $C/N$ and the chance of not drawing a corrupt
precinct,

\begin{align*}
     P(0) &= \frac{N-C}{N} \\
     \text{so that} \\
     P(0) &= 1-\frac{C}{N}.
\end{align*}

If we sample with replacement (each draw is an independent event) then the probability of drawing no corrupt precincts in $S$ draws is
\begin{equation*}
\left(1-\frac{C}{N} \right)^S
\end{equation*}
  and thus the probability of drawing one or more corrupt vote counts is
\begin{equation*}
  P=1-\left(1-\frac{C}{N} \right)^S.
\end{equation*}
If the rate of corrupt vote counts $c$ is $\frac{C}{S}$  where \(0 < c < 1\), then the chance of selecting zero corrupt counts in $S$ draws with replacement is \((1-c)^S\)  for selecting zero corrupt vote counts.  Therefore, the estimated probability for selecting one or more corrupt vote counts in S draws with replacement is  \(P=1-(1-c)^S\).

Beginning with the probability for not detecting any miscount
\begin{align*}
    1-P &\approx (1-c)^S \\
    \text{taking the log of both sides}\\
    \ln(1-P) &\approx \ln(1-c)^S \\
    \text{and}\\
    \ln(1-P) &\approx -cS \\
    \text{and solving for S gives}\\
    S &\approx \frac{\ln(1-P)}{-c}
\end{align*}

   To further improve the estimate for S, we use the formula for estimating the number of distinct elements, S, in a sample of size t drawn (with replacement) from a set of size N:\\
    \(S\approx N(1-e)^{\frac{-t}{N}}\)  and replace $t$ by our estimate for $S$ above, resulting in

\begin{equation}
S \approx N\left(1-e^\frac{\ln(1-P)}{Nc}\right)
\label{eq:S3}
\end{equation}
\cite{Rivest2007,Dopp2007a}.

Equivalently, because $Nc=C$, we get
\begin{align*}
    S &\approx N\left(1-e^\frac{\ln(1-P)}{C}\right)\\
      &\approx N\left(1-(e^{\ln(1-P)})^\frac{1}{C}\right)\\
      &\approx N\left(1-(1-P)^\frac{1}{C}\right),
\end{align*}
a simple formula for estimating post-election audit sample sizes to
provide at least $P$ probability for detecting one or more corrupt
vote counts in a sample of size $S$ if there are $N$ total vote counts
and $C$ miscounted vote counts.

  Then, as seen in Appendix~A, we can estimate
 \begin{align*}
   C &\approx \frac{N(w-r)}{k(b+w-r)},  \\
   S &\approx N\left(1-(1-P)^\frac{k(b+w-r)}{N(w-r)}\right).
 \end{align*}

A slightly more exact numerical method for calculating the
risk-limiting election audit sample size $S$ is found by solving the
sampling-without-replacement formula,  by employing the detailed estimate
for the minimum number of corrupt vote counts, $C$, using the gammaln
function for evaluating

\begin{equation*}
    \ln(1-P) - \ln[(N-C)!(N-S)!/(N!(N-C-S)!)],
\end{equation*}

\begin{eqnarray*}
     y&=&\ln(1-P)+\gammaln(N-C-S+1)-\gammaln(N-S+1)\\
      &&\qquad{}+\gammaln(N+1)-\gammaln(N-C+1).
\end{eqnarray*}

via the numerical method of bisections \cite{DoppStenger}.

\section*{Appendix~C: Double-Checking the Audit Sample Size}

Any post-election audit sample size S can be checked to see what minimum probability the sample size provides for detecting the minimum amount of vote miscount necessary to cause an incorrect election outcome by using the formula for the probability for drawing one or more miscounted audit units in a randomly drawn sample of audit units of size S, drawn without replacement, when there are $C$ corrupt vote counts out of a total of $N$ vote counts.
\begin{equation*}
 P=1-\frac{\binom{C}{0}\binom{N-C}{S}}{\binom{N}{S}}
 \end{equation*}
  can be calculated in a spreadsheet using the formula \texttt{1 -
  HYPGEOMDIST(0, S, C, N)} (See \cite{DoppBaiman}) This formula for
  checking the probability that the audit sample would detect
  outcome-altering vote miscount may be applied to both fixed rate
  audits and to risk-limiting audits.

\section*{Appendix~D: Some Weighted Random Sampling \& Error Bound Methods That Do Not Work Well}

This appendix points out details of the work of authors who have contributed new approaches to post-election auditing mathematics and methods that need some improvements.

For instance, the sampling weights recommended for post-election
auditing in \citetitle{Machine-Assisted Election Auditing} seem to be
incorrect \cite[pp. 7--8]{Calandrino2007a}.

Auditors desire a confidence level $c$ that no fraud significant
enough to change the election's outcome occurred. First the authors'
define their variables:

\begin{quotation}
``\dots let $v_1, \ldots , v_n$ be the electronically reported
vote totals for the candidates in decreasing order.
Therefore $v_1, \ldots , v_k$ correspond to winning candidates.
Because a single ballot may contain votes for up to $k$ candidates,
we need to consider the combination of votes on
each ballot.
	Given a ballot, let $C_s$, where $1 \leq s \leq k$, be the winning
candidate with the lowest vote total that received a
vote on the ballot. (Let $C_s$ be null if the ballot does not
contain votes for any winning candidate.) Let $C_t$, where
$k + 1 \leq  t \leq n$, be the losing candidate with the highest
vote total that did not receive a vote on the ballot. (Let
h$C_t$ be null if the ballot contains votes for all of the losing
candidates.)''
\end{quotation}

Calandrino et al.\ continue:

\begin{quotation}
	If $C_s$ is non-null, then we need to audit this ballot
	with probability at least $1-(1-c)^{1/b_1}$ , where $b_1 =
	v_s -v_{k+1}$. Intuitively, one possible result-changing
	scenario involving an error in this ballot would be
	to add $v_s - v_{k+1}$ incorrect votes for candidate $s$.
\end{quotation}

The last sentence of the preceding paragraph is incorrect because the most number of votes that any single ballot can possibly contribute is $k$ votes, where $k$ is the number of seats being elected in the contest. Thus the most number of incorrect votes that any one ballot can contribute is $k$ votes.

On the other hand, the difference between the initial reported number of votes for any winning and losing candidate could be in the thousands, a number much larger than the maximum number of incorrect votes $k$ that any ballot can produce in the initial reports.

In addition, the formula above cannot be generalized to audit units
with more than one ballot and does not seem to reliably produce
sensible sampling probabilities.%
\footnote{$p_i=1-(1-c)^{1/b_1}$.}

Similarly, the audit unit sampling weights recommended by Aslam, Popa,
and Rivest in \citetitle{On Auditing Elections When Precincts Have Different
Sizes} \cite[p. 16]{Aslam2008} described by the expression
\begin{quotation}
    $e_i = \min(2sv_i;M; v_i + r_{ij_1}-\min_{j}r_{ij})$ \\
    (It is OK just to use the first term, so that $e_i = 2sv_i$.)\ldots{}\\
    Also compute the total error bound:\\
    $E =\sum_{1\leq i\leq n}e_i$.
\end{quotation}
most often result in using the quantity $2sv_i$ where $s=0.20$ and
$v_i$ is the number of votes cast within each audit unit. The
quantity $2sv_i$ is then used for both the sampling weights and to calculate
the overall error $E$.  Using the quantity $2sv_i$ for weighting random selections is less desirable than weighting random selections by the number of ballots cast within each audit unit because $2sv_i$ does not account for under and over-votes since it uses the quantity ``votes cast'' rather than ``ballots cast''. Using $2sv_i$ rather than the actual upper margin error bounds will often result in an inadequate sample size and puts too much focus on auditing ballots that contain votes for losing candidates where not as much margin error could contribute to causing an incorrect initial outcome.

Figure~\ref{fig:Error-2svBound} below shows that the
quantity $2sv_i$ is most often less than the actual within audit unit
upper margin error bound, thus under-estimating
		the maximum possible margin error, increasing the quantity $M/E$ and
producing an inadequate sample size.  Figure~\ref{fig:Error-2svBound} also shows how $2sv$ is sometimes impossible because	there are not enough votes for the just-winning
		candidate in some audit units to contribute 40\%
		margin error. Note that the upper
		margin error bounds ``$b+w-r$'' and ``$2w+r+\text{other}$''
		can be as high as 200\% of the total number of within
		audit unit ballots while $2sv$ is always 40\% of the
		total number of ballots. Note that the actual upper
		margin error bounds increase as the winning candidate
		share increases, and that since the winning candidate
		normally has more within audit unit vote share than
		the losing candidate the $2sv$ error bound
		under-estimates the possible margin error, thus
		over-estimating the number of audit units needed to
		cause an incorrect outcome and under-estimating the
		post-election audit sample sizes necesssary to detect
		incorrect election outcomes.

 \begin{figure*}[hp] 
	\begin{center}
\includegraphics[width=0.9\textwidth]{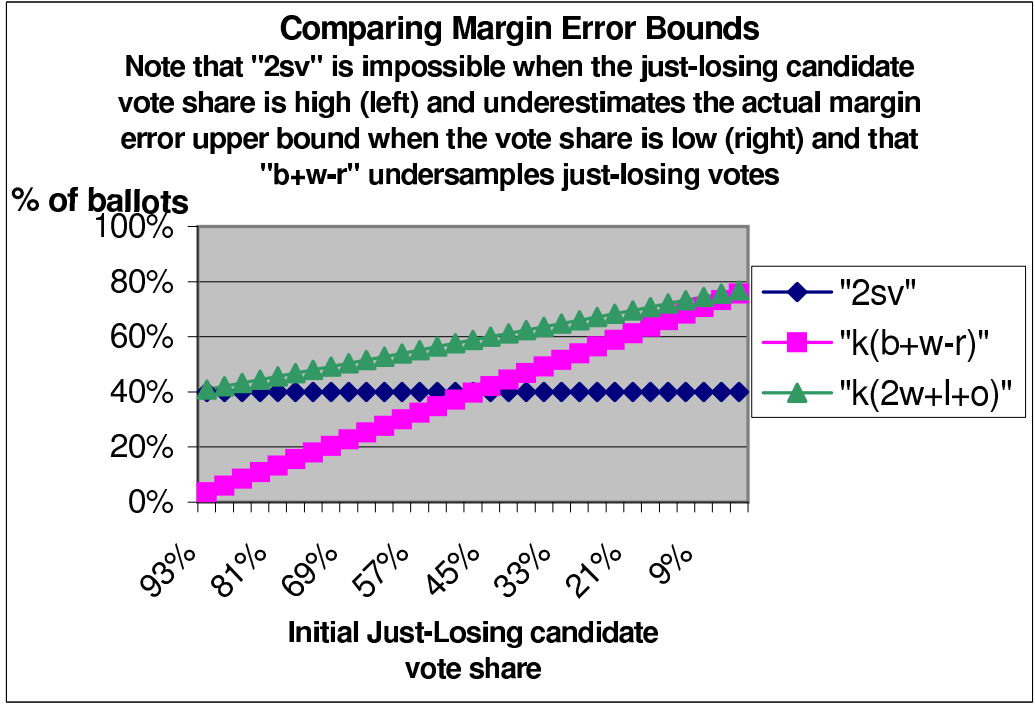}
	\caption[Error in $2sv$ bound]
		{The $2sv$ error bounds compared with actual upper
		margin error bounds for the ``just winning-losing''
		candidate pair, $k(b+w-r)$, and with the upper
		margin error bounds for \emph{all} ``winning-losing''
		candidate pairs, $k(2w+r+\text{other})$ where $k=2s$ ---
		plotted by initial vote share of the just-losing
		candidate}
	\end{center}
	\label{fig:Error-2svBound}
\end{figure*}

Despite being an excellent upper error bound for calculating sample sizes, the just-winning-losing candidate pair bounds, $b_i + w_{i}-r_{i}$, may be incorrect sampling weights.  As a sampling weight the just-winning-losing candidate pair bounds could increase the chance for certain election rigging strategies to prevail under some circumstances.

\section*{Appendix~E: The just-winning-losing candidate pair bounds produce the largest sample size} PROOF that using the winning-losing candidate pair with the smallest margin produces the most conservative (largest) post-election audit sample size:

Clearly $(w-r)\leq(w-l)$ where $l$ is the number of votes for any initial losing candidate.  So take any losing candidate with a greater margin than the runner-up and his margin can be expressed as $w-r+y$ where $y>0$.  Now compare the ratio $(M_r/E_r)$ of the runner-up and any other initial losing candidate.

We want to compare
\begin{equation*}
\frac{M_r}{E_r}=\frac{w-r}{b+w-r} \text{~and}
\end{equation*}
\begin{equation*}
\frac{M_l}{E_l}=\frac{w-(r-y)}{b+w-(r-y)}
\end{equation*}
to see which is bigger.

For simplicity, let $w-r = x$ so we compare
\begin{equation*}
\frac{M_r}{E_r}=\frac{x}{b+x} \text{~and}
\end{equation*}
\begin{equation*}
\frac{M_l}{E_l}=\frac{x+y}{b+x+y}
\end{equation*}
Multiplying both fractions top and bottom by the necessary factors to get a common denominator and comparing numerators we get

\begin{equation*}
xb+x_2+xy \leq xb+x_2+xy+by
\end{equation*}
 \begin{equation*}
 \text{~so that  }\frac{M_r}{E_r} \leq \frac{M_l}{E_l}
\end{equation*} Q.E.D.

\section*{Appendix~F: Example of maximum \emph{all} candidate-pair margin error bound}

If 200 ballots are cast in a precinct, and all four candidates --- two apparent winners and two apparent losers --- initially receive 50 votes apiece, the within audit unit upper margin error bound for \emph{all} winning-losing pairs is $2w + l = b + w = 2(100)+100 = 300$ votes where $w$ is the number of votes counted for initial winners and $l$ is the number of all other ballots, and $b$ is the number of total ballots cast.

This appendix shows one way out of the twelve possible ways that the maximum $300$ vote margin error can occur in this particular situation.

One way that the maximum $300$ vote margin error can occur is to assume that candidates A \& B are the initially, but incorrectly, reported winners, and that candidates C \& D are the initially incorrectly reported losers.  Given this scenario there are two possible scenarios that result in a 300 vote margin error.

Eg. Miscount the initial votes as follows in this precinct.
\begin{enumerate}
	\item	The 50 votes for candidate A should really have been counted for candidate C, causing a 100 vote margin error for the pair A \& C.
	\item	The 50 votes for candidate D should really have gone to candidate C, causing a 50 vote margin error for the pair A \& C.
	\item	The 50 votes counted for candidate B should really have been counted for candidate D, causing a 100 vote margin error for the pair B \& D.
	\item	The 50 votes counted for candidate C should really have been counted for candidate D, causing a 50 vote margin error for candidate pair B \& D.
\end{enumerate}

These errors add up to a 300 vote margin error for this scenario.

For each of the 6 possible pairs of winners and losers possible for this reviewer's scenario, there are 2 such examples of a 300 vote margin error, for 12 total examples of how a 300 vote margin error can occur.

\pagebreak

\section*{Appendix~G: Example of risk-limiting audit sampling methods in a close-margin election contest}

The tables below shows an example applying the margin error bounds and improved methods presented in this article to the election data for the close-margin 2004 Utah State Senate District \#~1 election contest.

\begin{table*}[ht] 
	\label{tab:EgDataMEBs2}
	\begin{widecenter}
	  \scriptsize
		\begin{tabular}[b]{|c|c|c|c|c|c|p{2.5cm}|p{2.5cm}|}
		\hline
		\multicolumn{8}{|c|}{\textbf{Example Upper Margin Error Bounds in a Narrow Margin Contest}} \\
		\hline
		\multicolumn{6}{|c|}{\textbf{2004 Utah State Senate, Dist \#~1 Vote Counts}} & \multicolumn{2}{|c|}{\textbf{Upper Margin Error Bnds}} \\
		\hline
		 {\textbf{Precinct}} & {\textbf{\#Ballots}} & {\textbf{Fife}} & {\textbf{Evans}} & {\textbf{Jenkins}}	&\textbf{} & {\textbf{Just-winning/just-losing candidate margin error bnds}} & {\textbf{All candidate pair margin error bnds}} \\
		 \hline
	\textbf{Totals}	&\textbf{16,987}	&	 \textbf{7,981} & {\textbf{7,553}}  & {\textbf{463}} & {\textbf{338}} & {\textbf{16,859}} & {\textbf{24,412}} \\
		\hline
	\textbf{SL2004}	&\textbf{574}	&  {\textbf{296}} & \textbf{233} & {\textbf{17}} & {\textbf{7}} & {\textbf{637}} & {\textbf{870}}\\
			\hline
	\textbf{SL2052}	&\textbf{518}	 & {\textbf{262}} & \textbf{191} & {\textbf{12}} & {\textbf{5}} & {\textbf{589}} & {\textbf{780}}\\
			\hline
	\textbf{SL1308}	&\textbf{536}	 & {\textbf{266}} & \textbf{226} & {\textbf{22}} & {\textbf{11}} & {\textbf{576}} & {\textbf{802}}\\
			\hline
	\textbf{SL2034}	&\textbf{572}	& \textbf{269} & {\textbf{275}} & {\textbf{8}} & {\textbf{4}} & {\textbf{566}} & {\textbf{841}}\\
			\hline
	\textbf{SL2226}	&\textbf{486}	& \textbf{260} & {\textbf{184}} & {\textbf{9}} & {\textbf{16}} & {\textbf{562}} & {\textbf{746}}\\
			\hline
	\textbf{SL2056}	&\textbf{518}	& \textbf{252} & {\textbf{212}} & {\textbf{23}} & {\textbf{9}} & {\textbf{558}} & {\textbf{770}}\\
			\hline
	\textbf{SL2006}	&\textbf{597}	& \textbf{247} & {\textbf{293}} & {\textbf{12}} & {\textbf{12}} & {\textbf{551}} & {\textbf{844}}\\
			\hline
	\textbf{SL2030}	&\textbf{545}	& \textbf{254} & {\textbf{260}} & {\textbf{8}} & {\textbf{13}} & {\textbf{539}} & {\textbf{799}}\\
			\hline
	\textbf{SL2214}	&\textbf{407}	& \textbf{238} & {\textbf{138}} & {\textbf{9}} & {\textbf{6}} & {\textbf{507}} & {\textbf{645}}\\
			\hline
	\textbf{SL2008}	&\textbf{453}	& \textbf{228} & {\textbf{186}} & {\textbf{16}} & {\textbf{10}} & {\textbf{495}} & {\textbf{681}}\\
			\hline
		\textbf{SL2224}	&\textbf{377}	& \textbf{237} & {\textbf{119}} & {\textbf{9}} & {\textbf{2}} & {\textbf{495}} & {\textbf{614}}\\
			\hline
		\textbf{SL1216}	&\textbf{499}	& \textbf{225} & {\textbf{232}} & {\textbf{13}} & {\textbf{12}} & {\textbf{492}} & {\textbf{724}}\\
			\hline
		\textbf{SL2049}	&\textbf{521}	& \textbf{224} & {\textbf{258}} & {\textbf{19}} & {\textbf{5}} & {\textbf{487}} & {\textbf{745}}\\
			\hline
		\textbf{SL1214}	&\textbf{562}	& \textbf{220} & {\textbf{300}} & {\textbf{12}} & {\textbf{8}} & {\textbf{482}} & {\textbf{782}}\\
			\hline
		\textbf{SL2242}	&\textbf{440}	& \textbf{219} & {\textbf{189}} & {\textbf{12}} & {\textbf{4}} & {\textbf{470}} & {\textbf{659}}\\
			\hline
		\textbf{SL2036}	&\textbf{517}	& \textbf{208} & {\textbf{272}} & {\textbf{9}} & {\textbf{10}} & {\textbf{453}} & {\textbf{725}}\\
			\hline
		\textbf{SL1306}	&\textbf{458}	& \textbf{199} & {\textbf{215}} & {\textbf{9}} & {\textbf{18}} & {\textbf{442}} & {\textbf{657}}\\
			\hline
		\textbf{SL2206}	&\textbf{333}	& \textbf{204} & {\textbf{102}} & {\textbf{9}} & {\textbf{13}} & {\textbf{435}} & {\textbf{537}}\\
			\hline
		\textbf{SL2014}	&\textbf{392}	& \textbf{206} & {\textbf{165}} & {\textbf{12}} & {\textbf{3}} & {\textbf{433}} & {\textbf{598}}\\
			\hline
		\textbf{SL2252}	&\textbf{346}	& \textbf{200} & {\textbf{116}} & {\textbf{9}} & {\textbf{5}} & {\textbf{430}} & {\textbf{546}}\\
			\hline
		\textbf{SL1302}	&\textbf{466}	& \textbf{193} & {\textbf{231}} & {\textbf{14}} & {\textbf{10}} & {\textbf{428}} & {\textbf{659}}\\
			\hline
		\textbf{SL1327}	&\textbf{455}	& \textbf{190} & {\textbf{227}} & {\textbf{14}} & {\textbf{10}} & {\textbf{418}} & {\textbf{645}}\\
			\hline
		\textbf{SL2003}	&\textbf{378}	& \textbf{198} & {\textbf{159}} & {\textbf{8}} & {\textbf{3}} & {\textbf{417}} & {\textbf{576}}\\
			\hline
		\textbf{SL2222}	&\textbf{383}	& \textbf{192} & {\textbf{159}} & {\textbf{7}} & {\textbf{10}} & {\textbf{416}} & {\textbf{575}}\\
			\hline
		\textbf{SL2246}	&\textbf{372}	& \textbf{194} & {\textbf{150}} & {\textbf{6}} & {\textbf{11}} & {\textbf{416}} & {\textbf{566}}\\
			\hline
		\textbf{SL1218}	&\textbf{374}	& \textbf{177} & {\textbf{147}} & {\textbf{12}} & {\textbf{21}} & {\textbf{404}} & {\textbf{551}}\\
			\hline
		\textbf{SL1328}	&\textbf{378}	& \textbf{182} & {\textbf{168}} & {\textbf{9}} & {\textbf{8}} & {\textbf{392}} & {\textbf{560}}\\
			\hline
		\textbf{SL2038}	&\textbf{290}	& \textbf{170} & {\textbf{86}} & {\textbf{14}} & {\textbf{1}} & {\textbf{374}} & {\textbf{460}}\\
		\hline
		\textbf{SL2254}	&\textbf{315}	& \textbf{167} & {\textbf{117}} & {\textbf{12}} & {\textbf{0}} & {\textbf{365}} & {\textbf{482}}\\
		\hline
		\textbf{SL2007}	&\textbf{454}	& \textbf{162} & {\textbf{267}} & {\textbf{7}} & {\textbf{8}} & {\textbf{349}} & {\textbf{616}}\\
		\hline
		\textbf{SL2002}	&\textbf{332}	& \textbf{160} & {\textbf{153}} & {\textbf{7}} & {\textbf{4}} & {\textbf{339}} & {\textbf{492}}\\
		\hline
		\textbf{SL1320}	&\textbf{411}	& \textbf{148} & {\textbf{229}} & {\textbf{13}} & {\textbf{7}} & {\textbf{330}} & {\textbf{559}}\\
		\hline
		\textbf{SL2204}	&\textbf{256}	& \textbf{148} & {\textbf{86}} & {\textbf{6}} & {\textbf{8}} & {\textbf{318}} & {\textbf{404}}\\
		\hline
		\textbf{SL1322}	&\textbf{321}	& \textbf{137} & {\textbf{142}} & {\textbf{15}} & {\textbf{11}} & {\textbf{316}} & {\textbf{458}}\\
		\hline
		\textbf{SL1350}	&\textbf{341}	& \textbf{129} & {\textbf{172}} & {\textbf{17}} & {\textbf{12}} & {\textbf{298}} & {\textbf{470}}\\
		\hline
		\textbf{SL1346}	&\textbf{347}	& \textbf{133} & {\textbf{190}} & {\textbf{8}} & {\textbf{8}} & {\textbf{290}} & {\textbf{480}}\\
		\hline
		\textbf{SL1210}	&\textbf{256}	& \textbf{116} & {\textbf{114}} & {\textbf{8}} & {\textbf{3}} & {\textbf{258}} & {\textbf{372}}\\
		\hline
		\textbf{SL1303}	&\textbf{343}	& \textbf{105} & {\textbf{199}} & {\textbf{7}} & {\textbf{9}} & {\textbf{249}} & {\textbf{448}}\\
		\hline
		\textbf{SL2050}	&\textbf{287}	& \textbf{104} & {\textbf{153}} & {\textbf{15}} & {\textbf{7}} & {\textbf{238}} & {\textbf{391}}\\
		\hline
		\textbf{SL2053}	&\textbf{196}	& \textbf{108} & {\textbf{67}} & {\textbf{6}} & {\textbf{3}} & {\textbf{237}} & {\textbf{304}}\\
		\hline
		\textbf{SL1351}	&\textbf{253}	& \textbf{85} & {\textbf{132}} & {\textbf{9}} & {\textbf{8}} & {\textbf{206}} & {\textbf{338}}\\
		\hline
		\textbf{SL2001}	&\textbf{73}	& \textbf{42} & {\textbf{25}} & {\textbf{0}} & {\textbf{1}} & {\textbf{90}} & {\textbf{115}}\\
		\hline
		\textbf{SL2216}	&\textbf{44}	& \textbf{27} & {\textbf{14}} & {\textbf{0}} & {\textbf{2}} & {\textbf{57}} & {\textbf{71}}\\
		\hline
		\end{tabular}
		\end{widecenter}
		\caption{The election data and margin error bounds for
		the 2004 Utah State Senate District \#~1 narrow-margin
		contest shown above are used in
		Tables~\ref{tab:Eg2UniformMethods}
		and~\ref{tab:EgCompareWeightedMethods2} to demonstrate
		the four improved methods for calculating
		risk-limiting post-election audit sample sizes
		presented in this paper.}
\end{table*}


\begin{table*}[tttt] 
	\label{tab:Eg2UniformMethods}
	\begin{widecenter}
	  \scriptsize
		\begin{tabular}[b]{|c|c|c|c|c|c|}
		\hline
		\multicolumn{6}{|c|}{\textbf{Example \#~2 of the Improved Uniform Method \& New Uniform Estimation Method}} \\
		\hline
		& \multicolumn{2}{|c|}{\textbf{Uniform Sampling Method}} & & \multicolumn{2}{|c|}{\textbf{Uniform Estimation Method}} \\
		\hline
		{\textbf{Sample Size}} & \multicolumn{2}{|c|}{\textbf{40}} & {|c|}{} &  \multicolumn{2}{|c|}{\textbf{$35 \approx S \leq 40$}} \\
		\hline
		 {\textbf{min \#corrupt AUs}} & \multicolumn{2}{|c|}{\textbf{2}} & {} & \multicolumn{2}{|c|}{\textbf{$2 \leq C \approx 3$}}  \\
			\hline
	\textbf{Precinct}	&	 \textbf{$ku_i$} & {\textbf{cumulative margin error}}  & \multicolumn{3}{|c}{}  \\
		\cline{1-3}
	\textbf{SL}	& \textbf{} & {\textbf{}} & \multicolumn{3}{|c}{}  \\
		\cline{1-3}
	\textbf{SL2004}	& \textbf{254.8} & {\textbf{254.8}} & \multicolumn{3}{|c}{}  \\
		\cline{1-3}
	\textbf{SL2052}	& \textbf{235.6} & {\textbf{490.4}} & \multicolumn{3}{|c}{}  \\
		\cline{1-3}
	\textbf{SL1308}	& \textbf{230.4} & {\textbf{}} & \multicolumn{3}{|c}{}  \\
		\cline{1-3}
	\textbf{SL2034}	& \textbf{226.4} & {\textbf{}} & \multicolumn{3}{|c}{}  \\
		\cline{1-3}
	\textbf{SL2226}	& \textbf{224.8} & {\textbf{}} & \multicolumn{3}{|c}{}  \\
		\cline{1-3}
	\textbf{SL2056}	& \textbf{223.2} & {\textbf{}} & \multicolumn{3}{|c}{}  \\
		\cline{1-3}
	\textbf{SL2006}	& \textbf{220.4} & {\textbf{}} & \multicolumn{3}{|c}{}  \\
		\cline{1-3}
	\textbf{SL2030}	& \textbf{215.6} & {\textbf{}} & \multicolumn{3}{|c}{}  \\
		\cline{1-3}
	\textbf{SL2214}	& \textbf{202.8} & {\textbf{}} & \multicolumn{3}{|c}{}  \\
		\cline{1-3}
	\textbf{SL2008}	& \textbf{198} & {\textbf{}} & \multicolumn{3}{|c}{}  \\
		\cline{1-3}
	\textbf{SL2224}	& \textbf{198} & {\textbf{}} & \multicolumn{3}{|c}{}  \\
		\cline{1-3}
	\textbf{SL1216}	& \textbf{196.8} & {\textbf{}} & \multicolumn{3}{|c}{}  \\
		\cline{1-3}
	\textbf{SL2049}	& \textbf{194.8} & {\textbf{}} & \multicolumn{3}{|c}{}  \\
		\cline{1-3}
	\textbf{SL1214}	& \textbf{192.8} & {\textbf{}} & \multicolumn{3}{|c}{}  \\
		\cline{1-3}
	\textbf{SL2242}	& \textbf{188} & {\textbf{}} & \multicolumn{3}{|c}{}  \\
		\cline{1-3}
	\textbf{SL2036}	& \textbf{181.2} & {\textbf{}} & \multicolumn{3}{|c}{}  \\
		\cline{1-3}
	\textbf{SL1306}	& \textbf{176.8} & {\textbf{}} & \multicolumn{3}{|c}{}  \\
		\cline{1-3}
	\textbf{SL2206}	& \textbf{174} & {\textbf{}} & \multicolumn{3}{|c}{}  \\
		\cline{1-3}
	\textbf{SL2014}	& \textbf{173.2} & {\textbf{}} & \multicolumn{3}{|c}{}  \\
		\cline{1-3}
	\textbf{SL2252}	& \textbf{172} & {\textbf{}} & \multicolumn{3}{|c}{}  \\
		\cline{1-3}
	\textbf{SL1302}	& \textbf{171.2} & {\textbf{}} & \multicolumn{3}{|c}{}  \\
		\cline{1-3}
	\textbf{SL1327}	& \textbf{167.2} & {\textbf{}} & \multicolumn{3}{|c}{}  \\
		\cline{1-3}
	\textbf{SL2003}	& \textbf{166.8} & {\textbf{}} & \multicolumn{3}{|c}{}  \\
		\cline{1-3}
	\textbf{SL2222}	& \textbf{166.4} & {\textbf{}} & \multicolumn{3}{|c}{}  \\
		\cline{1-3}
	\textbf{SL2246}	& \textbf{166.4} & {\textbf{}} & \multicolumn{3}{|c}{}  \\
		\cline{1-3}
	\textbf{SL1218}	& \textbf{161.6} & {\textbf{}} & \multicolumn{3}{|c}{}  \\
		\cline{1-3}
	\textbf{SL1328}	& \textbf{156.8} & {\textbf{}} & \multicolumn{3}{|c}{}  \\
		\cline{1-3}
	\textbf{SL2038}	& \textbf{149.6} & {\textbf{}} & \multicolumn{3}{|c}{}  \\
		\cline{1-3}
	\textbf{SL2254}	& \textbf{146} & {\textbf{}} & \multicolumn{3}{|c}{}  \\
		\cline{1-3}
	\textbf{SL2007}	& \textbf{139.6} & {\textbf{}} & \multicolumn{3}{|c}{}  \\
		\cline{1-3}
	\textbf{SL2002}	& \textbf{135.6} & {\textbf{}} & \multicolumn{3}{|c}{}  \\
		\cline{1-3}
	\textbf{SL1320}	& \textbf{132} & {\textbf{}} & \multicolumn{3}{|c}{}  \\
		\cline{1-3}
	\textbf{SL2204}	& \textbf{127.2} & {\textbf{}} & \multicolumn{3}{|c}{}  \\
		\cline{1-3}
	\textbf{SL1322}	& \textbf{126.4} & {\textbf{}} & \multicolumn{3}{|c}{}  \\
		\cline{1-3}
	\textbf{SL1350}	& \textbf{119.2} & {\textbf{}} & \multicolumn{3}{|c}{}  \\
		\cline{1-3}
	\textbf{SL1346}	& \textbf{116} & {\textbf{}} & \multicolumn{3}{|c}{}  \\
		\cline{1-3}
	\textbf{SL1210}	& \textbf{103.2} & {\textbf{}} & \multicolumn{3}{|c}{}  \\
		\cline{1-3}
	\textbf{SL1303}	& \textbf{99.6} & {\textbf{}} & \multicolumn{3}{|c}{}  \\
		\cline{1-3}
	\textbf{SL2050}	& \textbf{95.2} & {\textbf{}} & \multicolumn{3}{|c}{}  \\
		\cline{1-3}
	\textbf{SL2053}	& \textbf{94.8} & {\textbf{}} & \multicolumn{3}{|c}{}  \\
		\cline{1-3}
	\textbf{SL1351}	& \textbf{82.4} & {\textbf{}} & \multicolumn{3}{|c}{}  \\
		\cline{1-3}
	\textbf{SL2001}	& \textbf{36} & {\textbf{}} & \multicolumn{3}{|c}{}  \\
		\cline{1-3}
	\textbf{SL2216}	& \textbf{22.8} & {\textbf{}} & \multicolumn{3}{|c}{}  \\
		\cline{1-3}
	\textbf{SL2005}	& \textbf{0} & {\textbf{}} & \multicolumn{3}{|c}{}  \\
		\cline{1-3}

		\end{tabular}
		\end{widecenter}
		\caption{This table shows the improved and new uniform sampling calculations for the 2004 Utah State Senate District \#~1 narrow-margin contest (Data shown in Table~\ref{tab:EgDataMEBs2}). The confidence probability is $P = 0.99$ and the maximum level of undetectability is $k = 0.4$. The prior unimproved uniform method using $2sv_i$ error bounds calculates a significantly smaller sample size of 34 audit units for this contest because it under-est\-imates the within audit unit maximum margin error that could occur within each audit unit.}
\end{table*}

\begin{table*}[tttt] 
	\label{tab:EgCompareWeightedMethods2}
	\begin{widecenter}
	  \scriptsize
		\begin{tabular}[b]{|c|c|c|c|c|c|}
		\hline
		\multicolumn{6}{|c|}{\textbf{Example \#~2 applying the Improved Weighted Sampling Methods}} \\
		\hline
		& \multicolumn{2}{|c|}{\textbf{Improved PPMEB approach}} & & \multicolumn{2}{|c|}{\textbf{Improved PPMEBWR approach}} \\
		\hline
		{\textbf{Expected Sample Size}} & \multicolumn{2}{|c|}{\textbf{40}} & & \multicolumn{2}{|c|}{\textbf{34}} \\
		\hline
		 {\textbf{}} & {\textbf{}} & {\textbf{}} & {} & {\textbf{\#Draws}} & {\textbf{3}}  \\
		 \hline
	\textbf{Precinct}	&	 \textbf{$c_i$} & {\textbf{$p_i$}}  & {} & {\textbf{$p_i$}} & {\textbf{$1-(1-p)^t$}}  \\
		\hline
	\textbf{SL2004}	& \textbf{0.98} & {\textbf{0.99}} & {-} & {\textbf{0.03}} & {\textbf{0.92}} \\
		\hline
			\textbf{SL2052}	& \textbf{1.10} & {\textbf{0.98}} & {-} & {\textbf{0.03}} & {\textbf{0.90}} \\
		\hline
	\textbf{SL1308}	& \textbf{1.07} & {\textbf{0.99}} & {-} & {\textbf{0.03}} & {\textbf{0.91}} \\
		\hline
	\textbf{SL2034}	& \textbf{1.02} & {\textbf{0.99}} & {-} & {\textbf{0.03}} & {\textbf{0.92}} \\
		\hline
	\textbf{SL2226}	& \textbf{1.15} & {\textbf{0.98}} & {-} & {\textbf{0.03}} & {\textbf{0.89}} \\
		\hline
	\textbf{SL2056}	& \textbf{1.11} & {\textbf{0.98}} & {-} & {\textbf{0.03}} & {\textbf{0.90}} \\
		\hline
	\textbf{SL2006}	& \textbf{1.01} & {\textbf{0.99}} & {-} & {\textbf{0.03}} & {\textbf{0.92}} \\
		\hline
	\textbf{SL2030}	& \textbf{1.07} & {\textbf{0.99}} & {-} & {\textbf{0.03}} & {\textbf{0.91}} \\
		\hline
	\textbf{SL2214}	& \textbf{1.33} & {\textbf{0.97}} & {-} & {\textbf{0.03}} & {\textbf{0.85}} \\
		\hline
	\textbf{SL2008}	& \textbf{1.26} & {\textbf{0.97}} & {-} & {\textbf{0.03}} & {\textbf{0.87}} \\
		\hline
	\textbf{SL2224}	& \textbf{1.39} & {\textbf{0.96}} & {-} & {\textbf{0.02}} & {\textbf{0.84}} \\
		\hline
	\textbf{SL1216}	& \textbf{1.18} & {\textbf{0.98}} & {-} & {\textbf{0.03}} & {\textbf{0.88}} \\
		\hline
	\textbf{SL2049}	& \textbf{1.15} & {\textbf{0.98}} & {-} & {\textbf{0.03}} & {\textbf{0.89}} \\
		\hline
	\textbf{SL1214}	& \textbf{1.09} & {\textbf{0.99}} & {-} & {\textbf{0.03}} & {\textbf{0.90}} \\
		\hline
	\textbf{SL2242}	& \textbf{1.30} & {\textbf{0.97}} & {-} & {\textbf{0.03}} & {\textbf{0.86}} \\
		\hline
	\textbf{SL2036}	& \textbf{1.18} & {\textbf{0.98}} & {-} & {\textbf{0.03}} & {\textbf{0.88}} \\
		\hline
	\textbf{SL1306}	& \textbf{1.30} & {\textbf{0.97}} & {-} & {\textbf{0.03}} & {\textbf{0.86}} \\
		\hline
	\textbf{SL2206}	& \textbf{1.59} & {\textbf{0.94}} & {-} & {\textbf{0.02}} & {\textbf{0.80}} \\
		\hline
	\textbf{SL2014}	& \textbf{1.43} & {\textbf{0.96}} & {-} & {\textbf{0.02}} & {\textbf{0.83}} \\
		\hline
	\textbf{SL2252}	& \textbf{1.57} & {\textbf{0.95}} & {-} & {\textbf{0.02}} & {\textbf{0.80}} \\
		\hline
	\textbf{SL1302}	& \textbf{1.30} & {\textbf{0.97}} & {-} & {\textbf{0.03}} & {\textbf{0.86}} \\
		\hline
	\textbf{SL1327}	& \textbf{1.33} & {\textbf{0.97}} & {-} & {\textbf{0.03}} & {\textbf{0.85}} \\
		\hline
	\textbf{SL2003}	& \textbf{1.49} & {\textbf{0.95}} & {-} & {\textbf{0.02}} & {\textbf{0.82}} \\
		\hline
	\textbf{SL2222}	& \textbf{1.49} & {\textbf{0.95}} & {-} & {\textbf{0.02}} & {\textbf{0.82}} \\
		\hline
	\textbf{SL2246}	& \textbf{1.51} & {\textbf{0.95}} & {-} & {\textbf{0.02}} & {\textbf{0.81}} \\
		\hline
	\textbf{SL1218}	& \textbf{1.55} & {\textbf{0.95}} & {-} & {\textbf{0.02}} & {\textbf{0.80}} \\
		\hline
	\textbf{SL1328}	& \textbf{1.53} & {\textbf{0.95}} & {-} & {\textbf{0.02}} & {\textbf{0.81}} \\
		\hline
	\textbf{SL2038}	& \textbf{1.86} & {\textbf{0.92}} & {-} & {\textbf{0.02}} & {\textbf{0.74}} \\
		\hline
	\textbf{SL2254}	& \textbf{1.78} & {\textbf{0.93}} & {-} & {\textbf{0.02}} & {\textbf{0.76}} \\
		\hline
	\textbf{SL2007}	& \textbf{1.39} & {\textbf{0.96}} & {-} & {\textbf{0.02}} & {\textbf{0.84}} \\
		\hline
	\textbf{SL2002}	& \textbf{1.74} & {\textbf{0.93}} & {-} & {\textbf{0.02}} & {\textbf{0.77}} \\
		\hline
	\textbf{SL1320}	& \textbf{1.53} & {\textbf{0.95}} & {-} & {\textbf{0.02}} & {\textbf{0.81}} \\
		\hline
	\textbf{SL2204}	& \textbf{2.12} & {\textbf{0.89}} & {-} & {\textbf{0.02}} & {\textbf{0.70}} \\
		\hline
	\textbf{SL1322}	& \textbf{1.87} & {\textbf{0.91}} & {-} & {\textbf{0.02}} & {\textbf{0.74}} \\
		\hline
\textbf{SL1350}	& \textbf{1.82} & {\textbf{0.92}} & {-} & {\textbf{0.02}} & {\textbf{0.75}} \\
		\hline
\textbf{SL1346}	& \textbf{1.78} & {\textbf{0.92}} & {-} & {\textbf{0.02}} & {\textbf{0.76}} \\
		\hline
\textbf{SL1210}	& \textbf{2.30} & {\textbf{0.86}} & {-} & {\textbf{0.01}} & {\textbf{0.67}} \\
		\hline
\textbf{SL1303}	& \textbf{1.91} & {\textbf{0.91}} & {-} & {\textbf{0.02}} & {\textbf{0.73}} \\
		\hline
\textbf{SL2050}	& \textbf{2.19} & {\textbf{0.88}} & {-} & {\textbf{0.02}} & {\textbf{0.68}} \\
		\hline
\textbf{SL2053}	& \textbf{2.82} & {\textbf{0.81}} & {-} & {\textbf{0.01}} & {\textbf{0.59}} \\
		\hline
\textbf{SL1351}	& \textbf{2.53} & {\textbf{0.84}} & {-} & {\textbf{0.01}} & {\textbf{0.63}} \\
		\hline
\textbf{SL2001}	& \textbf{7.44} & {\textbf{0.46}} & {-} & {\textbf{0.00}} & {\textbf{0.29}} \\
		\hline
\textbf{SL2216}	& \textbf{12.06} & {\textbf{0.32}} & {-} & {\textbf{0.00}} & {\textbf{0.19}} \\
		\hline
		\end{tabular}
		\end{widecenter}
		\caption{This table shows two improved weighted sampling methods for risk-limiting post-election audits applied to the 2004 Utah State Senate District \#~1 narrow-margin contest (data shown in Table~\ref{tab:EgDataMEBs2}). This example uses a confidence probability of $P=0.99$ and a maximum level of undetectability of $k=0.4$. In this case, the improved PPMEBWR method shows an expected sample size of 34 audit units versus the old PPMEBR method using the $2sv$ error bound that would calculate a slightly smaller sample size of 33 audit units. The improved PPMEB method is more conservative than the improved PPMEBWR method with an expected sample size of 40 audit units.}
\end{table*}


\bibliographystyle{mod-authordate2}
\onecolumn
\bibliography{bibliography}

\end{document}